\magnification=1200 \baselineskip=13pt \hsize=16.5 true cm \vsize=20 true cm
\def\parG{\vskip 10pt} \font\bbold=cmbx10 scaled\magstep2

\centerline{{\it Braz. J. Phys.} {\bf 30}, 195 (2000)}\parG
\centerline{\bbold Broad Histogram: An Overview}\parG
\centerline{Paulo Murilo Castro de Oliveira}\parG
Instituto de F\'\i sica, Universidade Federal Fluminense\par
av. Litor\^anea s/n, Boa Viagem, Niter\'oi RJ, Brazil 24210-340\par
e-mail PMCO @ IF.UFF.BR\par

\vskip 0.5cm\leftskip=1cm\rightskip=1cm 

{\bf Abstract}\parG

        The Broad Histogram is a method allowing the direct
calculation of the energy degeneracy $g(E)$. This quantity is
independent of thermodynamic concepts such as thermal equilibrium.
It only depends on the distribution of allowed (micro) states
along the energy axis, but not on the energy changes between the
system and its environment. Once one has obtained $g(E)$, no further
effort is needed in order to consider different environment conditions,
for instance, different temperatures, for the same system.\par

        The method is based on the exact relation between $g(E)$ and
the microcanonical averages of certain macroscopic quantities $N^{\rm
up}$ and $N^{\rm dn}$. For an application to a particular problem,
one needs to choose an adequate instrument in order to determine the
averages $<N^{\rm up}(E)>$ and $<N^{\rm dn}(E)>$, as functions of
energy. Replacing the usual fixed-temperature canonical by the
fixed-energy microcanonical ensemble, new subtle concepts emerge. The
temperature, for instance, is no longer an external parameter
controlled by the user, all canonical averages being functions of
this parameter. Instead, the microcanonical temperature $T_{m}(E)$ is
a function of energy defined from $g(E)$ itself, being thus an {\bf
internal} (environment independent) characteristic of the system.
Accordingly, all microcanonical averages are functions of $E$.\par

        The present text is an overview of the method. Some features
of the microcanonical ensemble are also discussed, as well as some
clues towards the definition of efficient Monte Carlo microcanonical
sampling rules.\par

\leftskip=0pt\rightskip=0pt\parG


\vfill\eject

{\bf I -- Introduction}\parG

        The practical interest of equilibrium statistical physics is
to determine the canonical average

$$<Q>_T\,\, = \, {\sum_S Q_S \exp(-E_S/T) \over \sum_S
\exp(-E_S/T)}
\eqno(1)$$

\noindent of some macroscopic quantity $Q$. Here, the system is kept
under a fixed temperature $T$, and the Boltzmann constant is set to
unity. Both sums run over all possible microstates available for the
system, and $E_S$ ($Q_S$) is the value of its energy (quantity $Q$)
at the particular microstate $S$. The exponential Boltzmann factors
take into account the energy exchanges between the system and its
environment, in thermodynamic equilibrium.

        An alternative is to determine the microcanonical average

$$<Q(E)>\,\, = \, {\sum_{S[E]} Q_S \over g(E)}
\eqno(2)$$

\noindent of the same quantity $Q$. In this case, the energy is
fixed. Accordingly, the system is restricted only to the $g(E)$
degenerate microstates corresponding to the same energy level $E$,
and the sum runs over them.

        The canonical average (1) can also be expressed as

$$<Q>_T\,\, = \, {\sum_E <Q(E)> g(E) \exp(-E/T) \over \sum_E g(E)
\exp(-E/T)}\,\,\,\, ,
\eqno(3)$$

\noindent where now the sums run over all allowed energy levels $E$.

        Both $g(E)$ and $<Q(E)>$ depend only on the energy spectrum
of the system. They do not vary for different environment conditions.
For instance, by changing the temperature $T$, the canonical average
$<Q>_T$ varies, but not the energy functions $g(E)$ and $<Q(E)>$
which remain the same. Canonical Monte Carlo simulations are based on
equation (1), determining $<Q>_T$: one needs another computer run for
each new fixed value of $T$. Instead of this repetitive process, it
would be better to determine $g(E)$ and $<Q(E)>$ once and forever:
canonical averages can thus be calculated from (3), without
re-determining $g(E)$ and $<Q(E)>$ again for each new temperature.

        The Broad Histogram method [1] (hereafter, BHM) is based on
the exact relation (4), to be discussed later on. This equation
allows one to determine $g(E)$ from the knowledge of the
microcanonical averages $<N^{\rm up}(E)>$ and $<N^{\rm dn}(E)>$ of
certain macroscopic quantities: by fixing some energy jump
$\Delta{E}$, the number $N^{\rm up}_S$ ($N^{\rm dn}_S$) counts the
possible changes one could perform on the current microstate $S$,
yielding an energy increment (decrement) of $\Delta{E}$. Adopting
some microcanonical computer simulator in order to determine these
averages, and also $<Q(E)>$, one can calculate canonical averages
through (3) without further computer efforts for different
temperatures. As the Broad Histogram relation (4) is exact and
completely general for any system [2], the only possible source of
inaccuracies resides on the particular microcanonical simulator
chosen by the user.

        BHM does not belong to the class of reweighting methods
[3-7]. These are based on the energy probability distribution
measured from the actual distribution of visits to each energy level:
during the computer simulation, a visit counter $V(E)$ is updated to
$V(E)+1$ each time a new microstate is sampled with energy $E$. At
the end, the (normalized) histogram $V(E)$ measures the quoted energy
probability distribution. It depends on the particular dynamic rule
adopted in order to jump from one sampled microstate to the next. One
can, for instance, adopt a dynamic rule leading to canonical
equilibrium at some fixed temperature $T_0$. Then, the resulting
distribution can be used in order to infer the behaviour of the same
system under another temperature $T$ [3]. As canonical probability
distributions are very sharply peaked around the average energy,
other artificial dynamic rules can also be adopted in order to get
broader histograms [4-7], i.e. non-canonical forms of $V(E)$.

        BHM is completely distinct from these methods, because it
does not extract any information from $V(E)$. The histograms for
$N^{\rm up}(E)$ and $N^{\rm dn}(E)$ are updated to $N^{\rm up}(E)+
N^{\rm up}_S$ and $N^{\rm dn}(E)+N^{\rm dn}_S$ each time a new
microstate $S$ is sampled with energy $E$. Thus, the information
extracted from each sampled state $S$ is not contained in the mere
upgrade $V(E) \to V(E)+1$, but in the {\bf macroscopic} quantities
$N^{\rm up}_S$ and $N^{\rm dn}_S$ carrying a much more detailed
description of $S$. In this way, numerical accuracy is much higher
within BHM than any other method based on reweighting $V(E)$.
Moreover, the larger the system size, the stronger is this advantage,
due to the macroscopic character of $N^{\rm up}_S$ and $N^{\rm
dn}_S$.

        A second feature distinguishing BHM among all other methods
is its flexibility concerning the particular way one uses in order to
measure the fixed-$E$ averages $<N^{\rm up}(E)>$ and $<N^{\rm
dn}(E)>$. Any dynamic rule can be adopted in going from the current
sampled state to the next, provided it gives the correct
microcanonical averages at the end, i.e. a uniform visitation
probability to all states belonging to the {\bf same} energy level
$E$. The relative visitation frequency between different energy
levels $E$ and $E'$ does not matter. Any transition rate definition
from level $E$ to $E'$ can be chosen, provided it does not introduce
any bias {\bf inside each energy level, separately}. In this way, a
more adequate dynamics can be adopted for each different system,
still keeping always the full BHM formalism and definitions. On the
other hand, multicanonical approaches [5-7] are based on {\it a
priori} unknown transition rates: they are tuned during the
simulation in order to get a flat distribution of visits at the end,
i.e. a uniform $V(E)$ for which the visiting probability per state is
inversely proportional to $g(E)$. By measuring the actually
implemented transition rates from $E$ to $E'$, which has been tuned
during the computer run and must be proportional to $g(E)/g(E')$, one
can finally obtain $g(E)$. Thus, multicanonical approaches are
strongly dependent of the particular dynamic rule one adopts. Within
BHM, on the other hand, {\bf all} possible transitions between $E$
and $E'$ are {\bf exactly} taken into account by the quantities
$N^{\rm up}$ and $N^{\rm dn}$ themselves, not by the particular
dynamic transition rates adopted during the computer run.

        This text is divided as follows. Section II presents the
method, while some available microcanonical simulation approaches are
quoted in section III. In section IV, some particularities of the
microcanonical ensemble are discussed. Possible improvements in what
concerns microcanonical sampling rules are presented in section V.
Conclusions are in section VI.

\vskip 1cm
{\bf II -- The Method}\parG

        Consider a system with many degrees of freedom, denoting by
$S$ its current microstate. The replacement of $S$ by another
microstate $S'$ will be denoted {\bf a movement} in the space of
states. The first concept to be taken into account is the {\bf
protocol} of allowed movements. Each movement $S \to S'$ can be
considered allowed or forbiden, only, according to some previously
adopted protocol. Nothing to do with the probability of performing or
not this movement within some particular dynamic process: BHM is not
related to the particular dynamics actually implemented in order to
explore the system's space of states. We need to consider the {\bf
potential} movements which {\bf could} be performed, not the
particular path actually followed in the state space, during the
actual computer run. Mathematically, the protocol can be defined as a
matrix $P(S,S')$ whose elements are only 1 (allowed movement) or 0
(forbiden). The only restriction BHM needs is that this matrix must
be symmetric, i.e. $P(S,S') = P(S',S)$, corresponding to microscopic
reversibility: if $S \to S'$ is an allowed movement according to the
adopted protocol, so is the back movement $S' \to S$. Given a system,
one needs first to define this protocol. For an Ising magnet, for
instance, one can define that only single-spin flips will be
considered. Alternatively, one can accept to flip any set of spins up
to a certain number, or to flip clusters of neighbouring spins, or
any other protocol. BHM does not depend on which particular protocol
is adopted. This free choice can be used in order to improve the
efficiency of the method for each particular application (besides the
already quoted freedom in choosing the simulational dynamics, which
has nothing to do with this protocol of {\bf virtual} movements).

        Consider now all the $g(E)$ microstates $S$ belonging to the
energy level $E$ [8], and some energy jump $\Delta{E} > 0$ promoting
these states to another energy level $E' = E+\Delta{E}$. For each
$S$, given the previously adopted protocol, one can count the number
$N^{\rm up}_S$ [1] of allowed movements corresponding to this
particular energy jump. The total number of allowed movements between
energy levels $E$ and $E'$, according to the general definition (2)
of microcanonical average, is $g(E) <N^{\rm up}(E)>$. On the other
hand, one can consider all the $g(E')$ microstates $S'$ belonging to
the energy level $E'$ and the same energy jump $-\Delta{E} < 0$, now
in the reverse sense. For each $S'$ one can count the number $N^{\rm
dn}_{S'}$ [1] of allowed movements decreasing its energy by
$\Delta{E}$.  Due to the above quoted microscopic reversibility, the
total number $g(E') <N^{\rm dn}(E')>$ of allowed movements between
energy levels $E'$ and $E$ is the same as before. Thus, one can write

$$g(E) <N^{\rm up}(E)> \, = \, g(E+\Delta{E}) <N^{\rm dn}(E+\Delta{E})>
\,\,\,\, ,
\eqno(4)$$

\noindent which is the fundamental BHM equation introduced in [1].
The method consists in: a) to measure the microcanonical averages
$<N^{\rm up}(E)>$, $<N^{\rm dn}(E)>$ corresponding to a {\bf fixed}
energy jump $\Delta{E}$, and also $<Q(E)>$ for the particular
quantity $Q$ of interest, storing the results in $E$-histograms; b)
to use (4) in order to determine the function $g(E)$; and c) to
determine the canonical average $<Q>_T$ from (3), for any
temperature. Step a) could be performed by any means. Step b) depends
on the previous knowledge of, say, $g(0)$, the ground state
degeneracy. However, this common factor would cancel in step c).

        There is an alternative formulation of BHM [9], based on a
transition matrix approach [10]. Other alternatives can be found in
[11-14]. Interesting original analyses were presented in those
references. All of them differ from each other only on the particular
dynamics adopted in order to measure the BHM averages $<N^{\rm
up}(E)>$ and $<N^{\rm dn}(E)>$. The common feature is the BHM
equation (4). For multiparametric Hamiltonians, the energy $E$ can be
replaced by a vector $(E_1, E_2 \dots)$: in this way the whole phase
diagram in the multidimensional space of parameters can be obtained
from a single computer run [15], representing an enormous speed up.

        Besides the freedom of choosing the protocol of allowed movements,
the user has also the free choice of the energy jump $\Delta{E}$. In
principle, the same $g(E)$ could be re-determined again for different
values of $\Delta{E}$. Consider, for instance, the uniform Ising
ferromagnet on a $L \times L$ square lattice with periodic boundary
conditions, and only nearest-neighbour links, for an even $L > 2$. The
total energy can be computed as the number of unsatisfied links (pairs of
neighbouring spins pointing in opposed senses), i.e. $E = 0$, 4, 6, 8, 10
$\dots 2L^2-6$, $2L^2-4$, and $2L^2$ [16]. Adopting the single-spin-flip
protocol of movements, there are two possible values, namely $\Delta{E} =
2$ and $\Delta{E} = 4$, for the energy jumps. Thus, one can determine the
same $g(E)$ twice, within BHM. For that, one could store four distinct
$E$-histograms: $N^{\rm up}(\Delta{E}=4)$, $N^{\rm dn}(\Delta{E}=4)$,
$N^{\rm up}(\Delta{E}=2)$, $N^{\rm dn}(\Delta{E}=2)$. The first one
corresponds, for each microstate, to the number of spins surrounded by
four parallel neighbours, whereas the second to the number of spins
surrounded by four neighbours pointing in the opposed sense: together,
they determine $g(E)$ through (4), with $\Delta{E} = 4$, from the previous
knowledge of $g(0) = 2$ and $g(6) = 4L^2$. The third and fourth histograms
correspond, for each microstate, to the number of spins surrounded by just
three or just one parallel neighbours, respectively. Together, they can be
used, with $\Delta{E} = 2$, in order to determine $g(E)$ also through (4),
from the previous knowledge of $g(4) = 2L^2$.

        In practice, when Monte Carlo sampling is used as the
instrument to measure the microcanonical averages, this freedom on
the choice of $\Delta{E}$ can also be used in order to improve the
statistical accuracies [1,2,17-19], by taking all the possible values
of $\Delta{E}$ simultaneously. Provided one has always $\Delta{E} <<
E$ (hereafter the ground state energy will be considered as $E = 0$),
one can store only two $E$-histograms, with the combinations

$$N^{\rm up}_S = \sum_{\Delta{E}}\, [N^{\rm
up}_S(\Delta{E})]^{1/\Delta{E}}
\eqno(5{\rm a})$$

\noindent and

$$N^{\rm dn}_S = \sum_{\Delta{E}}\, [N^{\rm
dn}_S(\Delta{E})]^{1/\Delta{E}}
\,\,\,\, ,
\eqno(5{\rm b})$$

\noindent counted at each averaging state. This trick is an
approximation very useful in order to save both memory and time.
However, it could, in principle, introduce systematic errors,
depending on the particular application, and should be avoided in
those cases.

\vskip 1cm
{\bf III -- Some Microcanonical Simulation Approaches}\parG

        The averaged value defined by equation (2) refers to all
microstates $S[E]$ corresponding to the same energy level $E$, each
one being counted just once. For large systems, their number $g(E)$
is normally very large (except, in most cases, near the ground
state). In order to obtain a Monte Carlo (random sampling)
approximation for $<Q(E)>$, one actually averages only over a
restricted number of microstates, much smaller than $g(E)$. So, these
selected averaging states must represent the whole set without any
bias, besides the normal statistic fluctuations. The dynamic rule
must prescribe exactly the same sampling probability for each state.
The appropriate selection of these unbiased averaging states within
the same energy level is not an easy task. It depends on the
particular system under study, and also on the particular set of
allowed movements one adopts. There is no general criterion available
in order to assure the uniform sampling probability among fixed-$E$
microstates.

        One possible microcanonical simulation strategy is to perform
successive random movements, always keeping constant the energy. For
instance, the Q2R cellular automaton follows this strategy concerning
Ising-like models [20]. Each movement consists in: a) to choose some
spin at random; b) to verify whether the energy would remain the same
under the flipping of this particular spin; and c) to perform the
flip in this case. There is no proof that this strategy is unbiased.
Numerical evidences support this possibility, although good averages
are obtained only after very, very long transients [21]. On the other
hand, these enormous transient times can be avoided [22] by starting
the Q2R dynamics from a previously thermalized state, i.e. by running
first some canonical steps under a well chosen temperature
corresponding to the desired energy. One interpretation of these
findings is the following. Combined with the single-spin-flip
protocol, the fixed-energy dynamics is a very restrictive strategy in
what concerns a fast spread over the whole set of microstates with
energy $E$.  Indeed, numerical evidences of non ergodicity were found
[23]. Nevertheless, either by waiting enormous transient times or by
preparing the starting states, Q2R remains a possible choice for
microcanonical simulator of Ising systems. However, it will not give
good averages at all for very small energies.

        An alternative strategy is to relax a little bit the
fixed-energy constraint. This idea was introduced [24] by allowing
only small energy deviations along the path through the space of
states: successive random movements are accepted and performed,
provided they keep the energy always inside a small window, i.e.
always within a pre-defined set of few adjacent energy levels.
Although sampling different energy levels during the same run, the
visits to each one are taken into account separately by storing the
data in cummulative $E$-histograms. Even so, each energy level could
not be completely free of influences from the neighbouring ones.
Nevertheless, this strategy could be a good approximation provided:
a) the energy window width $\Delta{E}$ is very small compared with
the energy $E$ itself; and b) the final average $<Q(E)>$ is a smooth
function of $E$. Again, the method does not work very well for very
low energies, where the condition $\Delta{E} << E$ cannot be
fulfilled. This problem could be partially minimized by adopting some
smart tricks [25], although the very low energies would never be well
described [18].

        A third strategy is to relax completely the fixed-energy
constraint, by accepting any energy jump. Now, as $g(E)$ is a fast
increasing function of $E$, one is more likely to toss an energy
increment than a decrement. The result of accepting any tossed
movement would be a fast and irreversible arrival at the maximum
entropy region corresponding to infinite temperature, sampling
energies only near the maximum of $g(E)$. In order to avoid this, one
needs to introduce some acceptance restriction for energy-increasing
movements, trying to get a uniform sampling along the whole energy
axis. This was first introduced [26] within the distinct context of
finding optimal solutions (energy minima) in complex systems. The
idea is to divide all the possible movements one could perform,
starting from the current microstate, in two classes: increasing or
decreasing the energy. First, one of these two classes is randomly
tossed. Then, a random movement belonging to the tossed class is
performed. This strategy corresponds to an energy random walk, and
assures a uniform sampling along the whole energy axis, on average.
In spite of this very useful feature in what concerns the search for
optimum states in complex systems, this strategy cannot provide
correct thermal averages because different energy jumps are mixed
together, violating the relative Boltzmann weights between different
energy levels. In order to obtain correct thermal averages, one needs
to divide the possible movements not in two, but in as many classes
as different allowed energy jumps exist: for each fixed positive
$\Delta{E}$, one counts the number of increasing-energy movements ($E
\to E+\Delta{E}$) and the number of decreasing-energy ones ($E \to
E-\Delta{E}$), {\bf the same} $\Delta{E}$ for both, in order to
accept or not the currently tossed movement. In other words, one
needs to consider precisely the $\Delta{E}$-dependent BHM quantities
$N^{\rm up}_S$ and $N^{\rm dn}_S$ defined in [1].

        Many variants of this energy random walk dynamics can be
defined, the best one [27] being a direct consequence of the BHM
equation (4) itself, as follows. In order to obtain a flat
distribution of visits along the energy axis, one needs: a) to toss a
random movement starting from the current microstate with energy $E$;
b) to perform it whenever the energy decreases; and c) to perform it
only with probability $g(E)/g(E+\Delta{E})$, whenever the energy
increases ($\Delta{E} > 0$ being the increment). From the exact
relation (4), once one has some previous estimate of the energy
functions $<N^{\rm up}(E)>_0$ and $<N^{\rm dn}(E)>_0$, this
probability is also equal to

$$ p_{\rm up} =  
{<N^{\rm dn}(E+\Delta{E})>_0\over <N^{\rm up}(E)>_0}\,\,\,\, .
\eqno(6)$$

\noindent The dynamic rule proposed in [27] is then based on a
two-step computer simulation. First, one obtains an estimate of
$<N^{\rm up}(E)>_0$ and $<N^{\rm dn}(E)>_0$, by any means. Then,
using this first estimate in a further, independent computer run, one
performs the above defined dynamics, energy-increasing movements ($E
\to E+\Delta{E}$) being accepted only according to the probability
(6). From this second run, one measures accurate (according to [27])
averages $<N^{\rm up}(E)>$ and $<N^{\rm dn}(E)>$ from which $g(E)$
can be finally obtained from the BHM relation (4).

        This flat histogram dynamic recipe was proposed [27] in order
to solve some numerical deviations observed in previous versions
which could be viewed as approximations to it. As discussed in
section IV, unfortunately things are not so easy. These problems are
not merely related to particular dynamic approximate recipes, but to
another characteristic of the system itself: the discreteness of the
energy spectrum. All fundamental concepts leading to the
microcanonical ensemble are based on the supposition that all energy
changes $\Delta{E}$ are much smaller than the current energy $E$. In
other words, microcanonical ensemble is defined by disregarding the
energy spectrum discreteness. That is why conceptual problems appear:
a) at very low energies, for any system; b) at any energy, for tiny
systems. A better understanding of these subtle concepts cannot be
obtained by simple improvements of the dynamic recipe: a new
conceptual framework allowing to treat also discrete spectra is
needed (and lacking).

        On the other hand, excepting for the two situations quoted in
the above paragraph, the various approximations to the acceptance
rates (6) are not so bad as supposed in [27], according to the
evidences shown also in sections IV and V. Thus, let's quote these
approximations which make things easier. First, instead of performing
two computer runs in order to obtain the transition rates (6) from
the first and averages from the second, one can perform a single one
gradually accumulating $<N^{\rm up}(E)>$ and $<N^{\rm dn}(E)>$ in
$E$-histograms. At each step, in order to decide whether the
currently tossed movement must be performed or not, one uses the
already accumulated values themselves, by reading both the numerator
and the denominator of (6) from the corresponding histograms at the
proper energy channels $E$ and $E+\Delta{E}$. Actually, this trick
was already introduced (and used) in the original publication [1].
This approximation will be denoted by A1. It follows the same lines
of real-time-defined transition rates of the multicanonical sampling
methods [5-7]. A further approximation, hereafter called A2, consists
in ignoring the $\Delta{E}$ appearing in the numerator of (6),
reading both the numerator and the denominator from the same energy
channel $E$. This saves two real division operations by the current
number of visits $V(E)$ and $V(E+\Delta{E})$. The third additional
approximation consists in taking together all possible energy jumps
$\Delta{E}$ through equations (5a) and (5b), hereafter called A3.
This saves computer memory, because only a pair of histograms, one
for $<N^{\rm up}(E)>$ and the other for $<N^{\rm dn}(E)>$ are needed,
instead of a pair for each different possible value of $\Delta{E}$.
Also, this approximation saves time because the statistic
fluctuations are not spread over many histograms, the whole available
data being superimposed in only two.

        A further yet approximation is simply to replace the
numerator and the denominator of (6) by the current values $N^{\rm
up}_S$ and $N^{\rm dn}_S$ corresponding to the current microstate
$S$, instead of reading previously averaged values. In theory, for
large systems, this replacement could be justified if $N^{\rm up}$
and $N^{\rm dn}$ are shown to be self averaging quantities, in spite
of the further fluctuations it introduces. However, contrary to the
cases A1, A2 and A3 (described in the last paragraph and actually
tested [1,2,17-19]), this procedure does not save any computer time or
memory, being useless in practice.

        On the other hand, the procedure [28] of counting $N^{\rm
up}_S$ and $N^{\rm dn}_S$ at the current microstate $S$, {\bf without
classifying them according to different values of $\Delta{E}$}, is no
longer an approximation: it is wrong in what concerns the measurement
of averages, because the relative Boltzmann probability for different
energies would be violated. It corresponds to the mistake of missing
the exponent $1/\Delta{E}$ in equations (5a) and (5b). Only when
applied to the completely different context of finding optimal
solutions in complex systems [26], {\bf without performing averages},
this procedure is justifiable.

        The original multicanonical approach [5] can be re-formulated
according to the entropic sampling dynamics [6]. The degeneracy
function $g(E)$ is gradually constructed during the computer run. It
is based on an acceptance probability $g(E)/g(E')$ for each new
tossed movement from the current energy $E$ to another value $E'$.
After some steps, the whole function $g(E)$ is updated according to
the new visit trials, and so on. However, this method is yet based
exclusively on the histogram for $V(E)$. Another possibility [29] is
to adopt exactly this same dynamic rule in order to sample the
averaging states, measuring also the microcanonical averages $<N^{\rm
up}(E)>$ and $<N^{\rm dn}(E)>$ during the computer run. At the end,
$g(E)$ is obtained through the BHM equation (4), instead of the
values updated during the simulation: the results [29], of course,
are much more accurate. Note that, in this case, {\bf exactly} the
same microstates are visited, i.e. the same Markovian chain of
averaging states. In this way, the better BHM performance is
explicitly shown to be a consequence of the more detailed description
of each averaging state, compared with reweighting, multicanonical
approaches.

        Uniform probability distributions along the energy axis may
be not the best strategy. Being independent of the particular
simulation dynamics, BHM allows one to get better sampling statistics
in some energy regions. In [19] the Creutz dynamics [24,18] is
combined with the energy random walk [1]. First, one chooses a window
corresponding to few adjacent energy levels. Then, the energy random
walk dynamics is applied {\bf inside} this window up to a certain
predefined number of averaging states were sampled. After that, the
window is moved one energy level up, by including the next allowed
energy level at right and removing the leftmost. Then, the same
procedure is repeated inside this new window up to a slightly larger
number of averaging states, and so on. Reaching the critical region,
this number is kept fixed at its maximum value, before to go down
again for energies above the critical region. In this way, BHM allows
the user to design the profile of visits along the energy axis,
according to the numerical accuracy needed within each region.

\vskip 1cm
{\bf IV -- Canonical $\times$ Microcanonical Simulations}\parG

        The equivalence between the various thermodynamic ensembles
(canonical and microcanonical, for instance) is a widespread belief.
However, this is true only in the so-called thermodynamic limit $N
\to \infty$, where $N$ is the number of components forming the system
under study. For finite systems, and thus for any computer
simulational approach, the lack of this equivalence [30] poses
serious {\bf conceptual} as well as practical problems.

        Canonical computer simulation approaches were very well
developed since the pioneering work [31], half a century ago. By
following some precise recipes (random movements transforming the
current microstate into the next one) a Markovian chain of averaging
states is obtained, from which thermodynamic canonical averages are
calculated. As a consequence of this conceptual development,
particular recipes were shown to provide unbiased canonical
equilibrium. On the other hand, microcanonical simulation has never
attracted the attention of researchers during this same half century
(with some few exceptions). That is why some fundamental concepts
concerning this subject are misunderstood in the literature.

        The direct way to construct a fixed-$E$, microcanonical
simulator would be to accept a new randomly tossed movement only if
it does not change the energy. However, this constraint could
introduce non-ergodicity problems, depending on the particular set of
movements one adopts. For the Ising model, for instance, this problem
seems to occur by adopting only one-spin flips [21-23]. In order to
minimize it, one needs to allow more-than-one-spin flips [32].
However, by flipping only two spins far away from each other, after
each whole-lattice one-spin-flip sweep, the magnetic order is
destroyed below the critical energy [32]. Thus, the strictly fixed
energy approach is problematic, and the alternative is to relax it,
allowing some energy changes. However, this also poses troubles, once
the equilibrium features (magnetization, correlations, etc.) of the
system at some energy level $E$ are not exactly the same at another
level $E'$. By travelling from $E$ to $E'$ {\bf without time enough
for equilibration}, one could introduce biases from $E$-states into
$E'$ averages: the multiple-energy dynamics could distort the
strictly uniform probability distribution inside each energy level.
In short, the construction of a good microcanonical simulator is not
a simple subject.

        If you have not enough conceptual understanding about some
particular subject, a good idea is to resort to another similar
subject for which your conceptual understanding is already firmly
stablished.  Let's define a particular microcanonical simulator by
using well stablished canonical rules, for which the temperature $T$
is fixed since beginning, being an {\bf external control parameter}.
The canonical average of any macroscopic quantity $Q$ becomes a
function of $T$, as $<Q>_T$ in equations (1) or (3). Let's consider
some finite system with $N$ components, and its average energy
$<E>_T$.  Although the energy spectrum is discrete [8], $<E>_T$ is a
continuous function of $T$ (except for a possible isolated
temperature where a first order transition may occur). Thus, one can
tune the value of $T$ in order to have $<E>_T$ coincident with some
previously chosen energy level $E$: let's call this tuned temperature
$T(E)$. Then, a {\bf correct} microcanonical simulator recipe is: a)
to run some of the many known canonical recipes with fixed
temperature $T(E)$; b) to measure the quantity $Q_S$ of interest,
{\bf for each averaging state} $S$ {\bf whose energy is} $E$; c) to
accumulate $Q_S$ as well as the number of visits to the particular
energy level $E$, during the run; and d) to calculate the
microcanonical average $<Q(E)>$ at the end, by dividing the
accumulated sum of $Q_S$ by the number of visits to level $E$. This
recipe, of course, may not be very efficient, once one will visit
many energy levels, others than the previously fixed value $E$,
storing data concerning only this particular level. Also, the precise
temperature $T(E)$ must be known a priori, and perhaps this knowledge
could be achieved only by performing some previous runs. Moreover,
the whole process must be repeated for each different energy level.
Nevertheless, this recipe is perfectly correct, and will be the basis
for our reasonings hereafter.

        Table I shows the exact data for a square $4 \times 4$
lattice Ising ferromagnet, obtained by direct counting all the
$2^{16} = 65.536$ possible microstates. The adopted protocol of
(virtual) movements is the set of all possible one-spin flips. From
the first two columns, the exact temperatures $T(E)$ can be obtained
for each energy level $E$, by analytically calculating $<E>_T$ as a
function of $T$ from equation (3), and then imposing $<E>_T\, = E$.
Table II shows the data obtained from canonical simulation with fixed
$T(8) = 3.02866216$ (in the usual units where the energy
corresponding to each bond is $\pm J$, instead of 0 or 1), $10^8$
whole lattice sweeps, and $32$ independent samples. Thus, the total
number of averaging states is $3.2 \times 10^9$. For all averaged
values displayed, statistical fluctuations occur at most on the two
rightmost digits. The second column shows the actual number $V(E)$ of
visits to each energy level.  The expected statistical relative
deviations are, thus, of the order of $V(E)^{-1/2}$ in each case, in
agreement with the ones actually obtained (roughly 1 part in $10^4$).
Within this statistical accuracy, the coincidence between the first
line in Table II and the corresponding exact values for $E = 8$, in
Table I, is a further evidence confirming that the microcanonical
simulator defined in the above paragraph is indeed correct: it does
not introduce any bias besides the normal statistical numerical
fluctuations.

        Tables III to V show data obtained from canonical simulations
with fixed temperatures $T(10) = 3.57199419$, $T(12) = 4.66862103$
and $T(14) = 8.33883787$, respectively. These values were tuned in
order to give the exact average energies $<E>_T\, = 10$, 12 and 14,
respectively. Note again the coincidence of the second line in Table
III, the third line in Table IV, as well as the fourth line in Table
V with the exact values presented in Table I, within the numerical
accuracy.

        Level $E = 16$ is just the center of the energy spectrum,
corresponding to $T(16) = \infty$. In order to simulate this
situation, we adopted a very high temperature, namely $T = 100$, in
Table VI. Its fifth line is supposed to be compared with the
corresponding exact values in Table I.

        The other lines in Tables II to VI also coincide with the
corresponding exact values in Table I, within the numerical accuracy.
However, these further coincidences are not completely trustable, because
Table II, for instance, was obtained by fixing the simulational
temperature $T(8)$ tuned in order to give the average energy $<E>_T = 8$.
Thus, this Table II is out of tune for the other energy levels $E \neq 8$.
Accordingly, Tables III, IV, V and VI are out of tune for energies $E \neq
10$, $E \neq 12$, $E \neq 14$ and $E \neq 16$, respectively. Indeed,
considering for instance the simulations at fixed $T(E=14)$, the relative
deviation obtained for $N^{\rm dn}(E=8)$ with $\Delta{E} = 2$ and 4 are
respectively 14 or 18 times larger than the expected 1 part in $10^4$. In
principle, {\bf only} data obtained from canonical simulations performed
{\bf at the right temperature} $T(E)$ (i.e. the one for which $<E>_T\, =
E$) could be taken as microcanonical averages for this particular energy
level $E$. However, the systematic deviations induced by taking also data
corresponding to neighboring energy levels others than $E$ seem to be very
small. In particular, {\bf for larger systems, and near the critical
region}, justifying the simple approach of adding histograms obtained at
different temperatures [17,9]: in those cases, the temperature is a slowly
varying function of $E$ (${\rm d}T/{\rm d}E = 0$ at the critical point, in
the thermodynamic limit).

        In the thermodynamic limit, the canonical temperature can be
obtained by the statistical definition

$${1\over T} = {{\rm d}\over{\rm d}\epsilon}\,\,
\lim_{N\to\infty}\, {\ln g(N\epsilon)\over N}\,\,\,\, ,
\eqno(7)$$

\noindent where $\epsilon = E/N$ is the energy density, and coincides
perfectly with the value $T(E)$ quoted before, for which $<E>_T\, =
E$. However, for finite systems, both the thermodynamic limit
$N\to\infty$ as well as the derivative limit $\Delta{\epsilon} =
\Delta{E}/N \to 0$ cannot be performed. A paliative procedure is
simply to forget them, transforming equation (7) in

$${1\over{T_m(E)}} = {\ln g(E+\Delta{E}) - \ln g(E) \over
\Delta{E}}\,\,\,\, ,
\eqno(8{\rm a})$$

\noindent where the subscript $m$ means ``microcanonical
temperature'' defined just now for finite systems, and $\Delta{E}$ is
the energy gap between level $E$ and some other level above it. Of
course, in principle, $T_m(E)$ also depends on $\Delta{E}$, which
indicates a first difference between this and the true canonical
temperature. Moreover, contrary to the real, canonical temperature
$T(E)$, this $T_m(E)$ is not an external parameter depending on the
system's environment and equilibrium conditions: it is simply an
alternative formulation for the energy spectrum $g(E)$ itself, being
thus environment-independent. Note also that $T_m(E)$ is defined only
at the allowed energies belonging to the discrete spectrum, whereas
the real temperature $T(E)$ (the one for which $<E>_T\, = E$) can be
defined for any value $E$, continuously. An alternative formula is

$${1\over{T_m(E+\Delta{E}/2)}} = {\ln g(E+\Delta{E}) - \ln g(E) \over
\Delta{E}}\,\,\,\, .
\eqno(8{\rm b})$$

        The most famous canonical recipe [31] is: a) to toss some
random movement, starting from the current state; b) to calculate the
energy variation $\Delta{E}$ this movement would promote if actually
implemented; c) to perform the movement, whenever $\Delta{E} \leq 0$,
counting one more step; and d) to perform the movement only with the
Boltzmann acceptance probability $\exp(-\Delta{E}/T)$, if $\Delta{E}
> 0$, counting one step anyway. Normally, one takes a new averaging
state after $N$ successive steps (one MC step). Let's stress that,
here, the temperature $T$ is the real, canonical one: in order to use
this recipe to measure microcanonical averages at some fixed energy
$E$, one must take $T = T(E)$, the temperature for which $<E>_T\, =
E$. Instead, by taking $T = T_m(E)$ in equation (8a), it is an easy
exercise to show that the Boltzmann acceptance probability
$\exp(-\Delta{E}/T)$ would be equal to $g(E)/g(E+\Delta{E})$. But
this is just the acceptance probability adopted within the flat
histogram dynamics [27].

        Figure 1 shows $T_m(E)$ calculated from equation (8a), for a
$4 \times 4$ square Ising ferromagnet. The exact values of $g(E)$
were used. The open circles correspond to $\Delta{E} = 2$, and the
diamonds to $\Delta{E} = 4$. The continuous line is the exact $T(E)$,
also calculated from the exact values of $g(E)$. The deviations
between $T_m(E)$ and $T(E)$ are very strong. Note that there is no
approximation at all, neither in $T_m(E)$ nor in $T(E)$. The
deviations represent true differences between canonical and
microcanonical ensembles, which are indeed very strong for this tiny
system. Obviously, the condition $\Delta{E} << E$ is not fulfilled,
and the discreteness is inevitable along the whole energy spectrum.

        Figure 2 reports the same data as figure 1, with the same
symbols, now for a $32 \times 32$ lattice, and using equation (8b).
The exact values for $g(E)$ were taken from [33]. The same strong
deviations occur again, but now only at the very beginning of the
energy spectrum, where the condition $\Delta{E} << E$ does not hold,
as can be seen in the upper inset. However, near the critical region,
the deviations become much smaller, as can be seen in the lower inset
where the vertical scale is 100 times finer than the upper one. For
$E > 64$, the energy spectrum discreteness can be neglected within a
very good approximation, even with $N = 1024$ being still very far
from the thermodynamic limit in equation (7). The larger the system
size, the better becomes the situation, because the relation
$\Delta{E} << E$ becomes more and more fulfilled. However, at the
very beginning of the spectrum this relation will never be fulfilled,
even for large systems.

        Both limits in equation (7), namely the thermodynamic one $N
\to \infty$ and $\Delta{E} \to 0$ corresponding to the energy
derivative, were neglected in equation (8a). However, only the latter
seems to have disastrous consequences when one uses $T_m(E)$ as an
approximation for $T(E)$. Indeed, even for very small systems like
the $32 \times 32$ square lattice (not so tiny as $4 \times 4$), the
deviations are very small, provided the condition $\Delta{E} << E$
holds. In other words, the deviations between the microcanonical
temperature $T_m(E)$ and the true canonical value $T(E)$ comes almost
exclusively from the breakdown of the condition $\Delta{E} << E$,
{\bf not from finite size effects. Even in the thermodynamic limit,
$T_m(E)$ will differ from $T(E)$ near the ground state, due to the
discretness of the energy spectrum}. On the other hand, to take
$T_m(E)$ instead of $T(E)$ is a very, very good approximation far
from the ground state: we have seen before that microcanonical
averages obtained from canonical simulations are not sensitive to
temperatures out of tune for a given energy, moreover when the
deviations are of the order of that shown in the lower inset of
figure 2.

Tables II to VI present good numerical accuracies not only for the tuned
temperatures, but also when out-of-tune values for $T$ were considered.
Moreover, based on the reasonings above, this feature will be even
improved for larger and larger systems. The imediate consequence is the
possibility of adding different histograms for $<N^{\rm up}(E)>$ and
$<N^{\rm dn}(E)>$, obtained from distinct canonical simulations with
different temperatures, as already tested in [17,9]. Of course, this
approach is much more efficient than to take a different canonical
simulation with fixed temperature $T(E)$ for each different energy level
$E$, without superimposing the histograms. However, it still needs many
computer runs, one for each fixed temperature.

        In order to improve even more the efficiency, one can try the
following strategy. First, one defines the Boltzmann acceptance
probability $\exp[-\Delta{E}/T(E)]$ {\bf for each energy level $E$} and
each possible energy jump $\Delta{E}$. Note that this is not the same as
canonical simulations where the acceptance probability
$\exp(-\Delta{E}/T)$ depends only on $\Delta{E}$ but {\bf not on} $E$.
Then, by following this non-canonical, $E$-dependent acceptance
probability, one runs {\bf a single} computer simulation, visiting the
whole energy axis, accumulating the histograms for $<N^{\rm up}(E)>$ and
$<N^{\rm dn}(E)>$. This approach is similar to the flat histogram dynamics
[27], using the true temperature $T(E)$ instead of the microcanonical
value $T_m(E)$. Table VII shows the results for the same tiny system
already considered before. Surprisingly, strong deviations appear. In
order to analyse the source of these deviations, let's introduce another
very similar alternative, adopting a different acceptance probability
$\exp[-\Delta{E}/T(E+\Delta{E}/2)]$. The canonical Boltzmann probability
is taken {\bf at the center} $E+\Delta{E}/2$ of the interval correponding
to the energy jump from $E$ to $E+\Delta{E}$, not at the current energy
$E$. This symmetrization trick is supposed to diminish the numerical
deviations due to impossibility of performing the limit $\Delta{E} \to 0$,
i.e. the derivative in equation (7), for this tiny system. The results are
shown in Table VIII, in complete agreement with the exact results, Table
I. Thus, the source of the deviations in Table VII is the lack of the
condition $\Delta{E} << E$. By taking into account both the current energy
level $E$ as well as the (would-be) next energy $E+\Delta{E}$ {\bf in a
symmetric way}, the numerical deviations were eliminated. Of course, for
larger systems and far from the ground state, where the condition
$\Delta{E} << E$ holds, the differences between Tables VII and VIII would
also disappear.

        The flat histogram dynamics uses another acceptance probability,
namely

$$\exp[-\Delta{E}/T_m(E)] = {g(E)\over g(E+\Delta{E})} = {<N^{\rm
dn}(E+\Delta{E})>\over <N^{\rm up}(E)>}\,\,\,\, ,
\eqno(9)$$

\noindent where $E$ and $E+\Delta{E}$ also play a {\bf symmetric} role.
Simulational results are shown in Table IX, where the exact values for
$g(E)$ (or, alternatively, $<N_{\rm up}(E)>$ and $<N_{\rm dn}(E)>$) were
adopted in order to determine the acceptance probabilities (9). The
results are again coincident with the expected ones, Table I [34]. On the
other hand, if one tries to break the quoted symmetry, ignoring
$\Delta{E}$ in the numerator of the right-hand-side term in (9), i.e.
approximation A2, numerical deviations similar to Table VII would appear,
for this tiny system.

        As one does not know a priori $g(E)$ (or, alternatively, $<N_{\rm
up}(E)>$ and $<N_{\rm dn}(E)>$), one can use some previous estimates
$<N_{\rm up}(E)>_0$ and $<N_{\rm dn}(E)>_0$, and adopts the acceptance
probability (6). In order to follow this recipe [27], one needs to
determine the quoted estimates from a previous computer run. According to
[27], these previous estimates do not need to be very accurate. However,
actual numerical tests [29] show results for $<N_{\rm up}(E)>$ and
$<N_{\rm dn}(E)>$ worse than the inputs $<N_{\rm up}(E)>_0$ and $<N_{\rm
dn}(E)>_0$ themselves! A better possibility is the random walk dynamics
originally used in order to test the broad histogram method [1]. It is the
same as the flat histogram, equation (6) or (9), with the approximation A1
(optionally, also A2 and A3) quoted in last section. A1 consists in taking
the current, already accumulated values of the histograms $N^{\rm up}(E)$
and $N^{\rm dn}(E+\Delta{E})$, in real time during the computer run
itself, instead of the true averages at the right-hand side of equation
(6) or (9). Results for the same tiny system treated before are shown in
Table X. This approach is the same as the multicanonical
real-time-measured transition probability already adopted in other earlier
methods, for instance the entropic sampling [6]. A criticism to this
approach is its non-strictly-Markovian, history-dependent character.
According to [29], it is actually better than the fixed transition
probability proposed in [27]. Approximation A2 consists in neglecting
$\Delta{E}$ in the numerators of equation (6) or (9), breaking the
symmetry between levels $E$ and $E+\Delta{E}$, and cannot be applied to
this tiny system due to the spectrum discretness. This approach A2 is also
shown to violate a particular detailed balance condition [27]. A3 consists
in taking all possible values for $\Delta{E}$ together, by using equations
(5a) and (5b). Both approximations A2 and A3 (but not A1) are bad: a) at
very low energies, for any system; b) at any energy, for tiny systems.

        The difference between the dynamics adopted in Tables VII to X and
the true canonical rule adopted in Tables II to VI is the following.
Canonical simulations adopt {\bf the same} acceptance probability
$\exp(-\Delta{E}/T)$ for any tossed movement increasing the energy by
$\Delta{E}$, no matter which is the current energy $E$. As a consequence,
the energies visited during the run became restricted to a narrow window
around the canonical average $<E>_T$. The larger the system size, the
narrower is this window. On the other hand, within the non-canonical rules
adopted in Table VII to X, the acceptance probabilities depend also on $E$
(and $E+\Delta{E}$). As a consequence, instead of a narrow distribution of
visits, one gets a broad distribution covering the whole energy axis. This
feature is, of course, a big advantage over canonical rules, because only
one run would be enough to cover a large temperature range. Nevertheless,
the numerical results could be wrong, depending on the actual dynamic rule
one adopts (Table VII, for instance). A better conceptual, theoretical
understanding of these and other $E$-dependent dynamic rules is needed,
and lacking. Concerning BHM, the only constraint to be considered is the
uniform probability visitation inside each energy level, separately. For
other, reweighting methods based on the actual visitation profile $V(E)$,
also the detailed relative distribution between different energy levels
must be considered.

        Up to now, we have tested many different dynamic rules in order to
measure the microcanonical averages $<N_{\rm up}(E)>$ and $<N_{\rm
dn}(E)>$ from which one can determine the desired quantity $g(E)$ by BHM
equation (4). The most efficient approaches are the $E$-dependent rules
(Tables VIII, IX and X), where a single computer run is enough. Among
them, under a practical point of view, the random walk dynamics [1]
corresponding to Table X is the best choice, once one does not need any
previous knowledge about the quantities $<N_{\rm up}(E)>$ and $<N_{\rm
dn}(E)>$ to be measured. This approach corresponds to approximation A1.
The other two further approximations A2 and A3 [1] improve even more the
efficiency. However, due to the energy spectrum discreteness, they are
limited by the constraint $\Delta{E} << E$: one needs to avoid them for
tiny systems, or very near the ground state even for large systems, where
this constraint cannot be fulfilled. All this matter corresponds to the
subject discussed in reference [27]. Let's now discuss another, subtle,
further possible source of unaccuracies, which may appear when one abandon
the safe canonical dynamical rules and adopts non-canonical, $E$-dependent
dynamics in order to sample the whole energy axis during the same computer
run.

        In order to introduce the subject, let's resort again to canonical
simulations, where some temperature value $T$ is fixed since beginning.
Imagine one starts such a simulational process from a randomly chosen
microstate $S$: its energy $E_S$ would be in general far from the average
value $<E>_T$, as well as many other features of this microstate which
would be far from their equilibrium counterparts. By plotting the energy
of each successive averaging microstate as a function of the time, one
would get a curve fluctuating around an exponential decay to the final
value $<E>_T$: at the end, only statistical fluctuations around this
constant value remain. Thus, the very beginning of the Markovian chain of
states will give wrong (biased) contributions to the final averages
$<Q>_T$. These are the so-called thermalization transient steps.
Nevertheless, they represent no problem at all, because one can always
take an enormous number of microstates {\bf after} this transient bias is
already over, pushing the systematic deviations to below any predetermined
tolerance. Better yet, one can simply discard the contribution of these
initial states, by starting the averaging procedure only after the
transient is over. In the statistical physicists' jargon, one can assert
that ``the system is already thermalized'', after these initial
out-of-equilibrium, transient steps.

        As quoted in [27], detailed balance is a delicate matter. Detailed
balance conditions are useful only in order to ensure that the final
distribution, chosen by the user, will be the correct one, for instance the
Gibbs distribution for equilibrium canonical averages. However, these
conditions do not ensure this final distribution will be reached within a
{\bf finite} time. In our case of interest, i.e. the $E$-dependent,
broad-energy, non-canonical dynamics, the system {\bf never} reaches
canonical equilibrium inside each energy level. All these dynamic rules
correspond to generalized Boltzmann factors $\exp[-\Delta{E}/\Theta(E)]$,
where the ``temperature'' $\Theta(E)$ varies from one microstate to the
next, along the Markovian chain. Concerning the energy, for instance,
instead of an exponential decay to the canonical equilibrium situation,
one gets a random walk visiting {\bf all} energies, with fluctuations
covering {\bf the whole} energy axis {\bf all} the time. The same large
fluctuation behaviour holds also for other quantities, in particular the
one for which the microcanonical averaging is in progress. This dynamics
follows an {\bf eternal transient} in what concerns the safe canonical
framework. This feature may introduce systematic numerical deviations.
Although those possible out-of-canonical-equilibrium problems cannot be
observed in our Tables VIII, IX and X, they could be crucial for larger
systems where large energy jumps in few steps become possible: features of
a particular far energy region which the system recently cames from could
introduce biases in the current energy averages. In this eternal-transient
case, detailed balance conditions and all the related theorems give little
help. To construct a good, efficient microcanonical simulator, by visiting
the whole energy axis during a single computer run, seems to be a much
more delicate matter. How to assure a uniform probability distribution
{\bf inside each energy level} also covering broad regions of the energy
spectrum? This problem is open to new insight, new ideas.

        Exemplifying how difficult would be to analyse the uniformity of
visits within each energy level, let's take a simple example: a $L \times
L$ square lattice Ising magnet ($L > 4$, $N = L^2$ spins), with $E = 8$.
Considering the magnetization density $m$, level $E = 8$ is divided into
three classes. The first one contains $g_1(8) = N(N-5)$ states with only
two non-neighbouring spins pointing down (all other spins up, or
vice-versa), corresponding to $m = 1 - 4/N$. The second class consists of
$g_2(8) = 12N$ states with a 3-site cluster of neighbouring spins pointing
down, with $m = 1 - 6/N$. The third class has $g_3(8) = 2N$ states
presenting a plaquette of four spins pointing down, and $m = 1 - 8/N$. No
single-spin flip can transform a state with $E = 4$ or $E = 6$ into
another state belonging to this third class, for instance. In order to
assure a uniform distribution of visits, this lack must be compensated by
other possible single-spin flips from $E = 10$ and $E = 12$. On the other
hand, the visitation frequency to each one of these higher-energy states
depends on feeding rates from higher yet energies, and so on. One must
prescribe adequate transition probabilities to each such movement, taking
into account all its consequences on the next, next-next step, and so on.
Chessboard is an easier game. Fortunately, the first class containing
$\sim N^2$ states dominates the counting for large lattice sizes. For $L =
32$, for instance, $g_1(8) = 1,043,456$ states are in the first class,
representing $99\%$ of the whole number $g(8) = g_1(8) + g_2(8) + g_3(8) =
1,057,792$. Similar behaviours also occur for higher energies, giving us a
solace: we remain with the hope that single-spin flips may lead to the
desired uniformity for large enough systems. For a tiny $4 \times 4$
lattice, on the other hand, things go worse, once two further classes must
be added to $E = 8$ level: $g_4(8) = 16$ states with a single line of
spins down, corresponding to $m = 1/2$; and $g_5(8) = 8$ states with two
neighbouring lines of spins down, with $m = 0$. Then, only $g_1(8) = 176$
states out of $g(8) = 424$, i.e. $42\%$, belong to the first class. This
feature, of course, partially explains the bad results in Table VII.
However, a detailed explanation for the good results obtained in Tables
VIII, IX and X, following the same reasonings, is not easy.

\vskip 1cm
{\bf V -- Improved Microcanonical Simulators}\parG

        While we have not yet a perfect and efficient microcanonical
simulator, the advantages of the Broad Histogram method, i.e. the
exact and completely general equation (4), are already in hands. The
only problem is to obtain good estimates for the averages $<N^{\rm
up}(E)>$ and $<N^{\rm dn}(E)>$, as functions of $E$. One paliative
solution is to use the available, not-yet perfect microcanonical
simulators, perhaps with some repairing procedures.  Even without any
repair, the random walk naive approach introduced in [1] can be
useful. First, observing Tables II to VI, one notes that some (large)
degree of inaccuracy on the temperature can be tolerated still within
very good precision on the final results. Second, comparing figures 1
and 2, one can note that larger systems present a plateau of almost
constant temperatures covering a large energy range. This range is
just the interesting one, namely the critical region where the
specific heat diverges for $N \to \infty$. Third, with small
variations of the temperature from one microstate to the next, within
this region, the out-of-canonical-equilibrium problems should be
strongly diminished.

        Anyway, we can try to introduce some repairing procedures into the
random walk dynamics [1] (or into any other $E$-dependent transition rate
dynamics). The first idea is to equilibrate the system before measuring
any averaging quantity at the current microstate. This can be acomplished
by running some canonical steps just before the measuring procedure.
Suppose one gets some microstate with energy $E$, during the random walk
dynamics. As this microstate comes from other energy levels, submitted to
acceptance probabilities others than the currently correct value
$\exp[-\Delta{E}/T(E)]$, it is supposed to carry some undesired biases.
Then, one can simply include some canonical steps with fixed temperature
$T = T(E)$, in order to let the system relax to $E$-equilibrium, before
taking the averages. $T(E)$ can be estimated at each step from the
current, already accumulated histograms for $N^{\rm up}(E)$ and $N^{\rm
dn}(E)$ (approximation A1), and the further approximations A2 and A3 can
also be adopted. All these tricks were already introduced in the original
publication [1].

        Another repairing procedure is simply to forbid large energy
jumps. The simplest way to perform this task is to count a new averaging
state after each single-spin-flip trial. Normally, one adopts $N$ trials,
i.e. a whole lattice sweep, before counting a new averaging state, in
order to avoid possible correlations along the Markovian chain of states.
In our microcanonical case, however, different energy levels correspond to
independent averaging processes, and one can try to abandon this
precaution. Of course, this approach also saves a lot of computer time.
Table XI shows the results obtained for a $32 \times 32$ lattice, where
approximation A1 (real-time measurement of the transition rates) was
adopted, with $10^9$ averaging states per energy level. The expected
relative error due to finite statistics is thus $3 \times 10^{-5}$.
Indeed, the observed deviations coincide with this, in spite of the figure
$0.008$ [27] predicted by detailed balance arguments. Even adopting the
further approximations A2 (which explicitly violates detailed balance) and
A3, the errors remain the same (last column). Once more, the possible
source of unaccuracies has nothing to do with detailed balance dictating
the relative frequency of visits between {\bf different energy levels}. On
the contrary, the only crucial point is the uniformity of visits {\bf
inside each energy level, separately}.

        Perhaps the procedure of measuring averages after each new
single-spin-flip trial could indeed introduce some bias, through some
undesirable correlations along the Markovian chain, although this
possibility is not apparent at all in Table XI. Perhaps this bias
could appear only for much larger statistics. An alternative way to
avoid large energy jumps still taking averages only after each whole
lattice sweep is to restrict the random walk to narrow energy
windows. This corresponds to the Creutz energy-bag simulator [24]
combined with the random walk dynamics [1] inside each window.  It
was introduced in [19], and the results also show relative errors
much smaller than those predicted by detailed balance arguments. The
specific heat for a $32 \times 32$ square lattice is shown in figure
3, where approximations A1, A2 and A3 were used. By restricting the
energy to narrow windows, one is also restricting the temperature to
small variations, thus forcing the system to be always near the
canonical equilibrium conditions for any energy inside the current
window. Once again, the best performance occurs at the critical
region.

\vskip 1cm
{\bf VI -- Conclusions}\parG

        The Broad Histogram Method (BHM) introduced in [1] allows one
to determine the energy spectrum of any system, i.e. the degeneracy
$g(E)$ as a function of the energy $E$, through the {\bf exact and
completely general} equation (4). First, one needs to adopt some
reversible protocol of allowed movements in the system's space of
states.  Reversible means that for any allowed movement $S \to S'$,
the back movement $S'\to S$ is also allowed. For Ising models, for
instance, one can choose single-spin flips, to flip clusters of
neighboring spins, etc. In fact, one could invent {\bf any} protocol.
For each state $S$, $N^{\rm up}_S$ counts the number of such allowed
movements for which the system's energy would be increased by a {\bf
fixed} amount $\Delta{E}$. Accordingly, $N^{\rm dn}_S$ counts the
number of allowed movements decreasing the energy by {\bf the same}
amount $\Delta{E}$. The energy jump $\Delta{E}$ is also chosen and
fixed since the beginning. $<N^{\rm up}(E)>$ and $<N^{\rm dn}(E)>$
are the microcanonical, fixed-$E$ averages for these two quantities,
both functions of the energy. Then, in order to determine $g(E)$ from
the BHM equation (4), one needs only to find a way, any way, to
measure those microcanonical averages.

        By following the same way adopted in measuring $<N^{\rm
up}(E)>$ and $<N^{\rm dn}(E)>$, one can also obtain the
microcanonical average $<Q(E)>$ of the particular thermodynamic
quantity $Q$ of interest (magnetization, density, correlations, etc).
{\bf All} these microcanonical averages are {\bf independent} of the
particular environment the system is currently interacting with. In
other words, they do not depend on temperatures, equilibrium
conditions, or any other thermodynamic concept: they are determined
by the system's energy spectrum alone. Thus, after $g(E)$ is already
determined through the BHM equation (4), once and forever, the same
system can be submitted to different environment conditions, and its
behaviour (equilibrium or not) can be studied resorting always to the
same $g(E)$. Within the particular case of canonical equilibrium, for
instance, the thermal average $<Q>_T$ of the quantity $Q$ can be
obtained from equation (3), for any temperature $T$. If one decides
to use computer simulations as the instrument measuring $<N^{\rm
up}(E)>$, $<N^{\rm dn}(E)>$ and $<Q(E)>$, then only one computer run
is enough to determine the whole temperature dependence,
continuously, without need of repeating again the simulation for each
new temperature.

        Other computer simulation methods, the so-called
multicanonical sampling strategies [5-7], also allow the direct
determination of $g(E)$. All of them are based on the counting $V(E)$
of visits to each energy level $E$. Every time each energy level $E$
is visited by the current state $S$, along the Markovian chain, the
counting is updated from the current $V(E)$ to $V(E) + 1$. Within
BHM, however, the $E$-histograms for $N^{\rm up}$ and $N^{\rm dn}$
are updated from the current $N^{\rm up}(E)$ and $N^{\rm dn}(E)$ to
$N^{\rm up}(E) + N^{\rm up}_S$ and $N^{\rm dn}(E) + N^{\rm dn}_S$,
respectively, for the same current state $S$. This corresponds to
{\bf macroscopic} upgrades instead of the mere counting of one more
state. Thus, a much more detailed information is extracted from each
averaging state withing BHM than any other method, giving rise to
much more accurate results. Moreover, the larger the system size, the
stronger is this advantage, due to the macroscopic character of the
BHM quantities $N^{\rm up}$ and $N^{\rm dn}$.

        Another advantage of BHM over other methods is its complete
independence concerning the particular dynamic rule adopted in order
to measure the microcanonical averages. {\bf Any} recipe can be used,
provided the correct $E$-functions $<N^{\rm up}(E)>$, $<N^{\rm
dn}(E)>$ and $<Q(E)>$ were accurately determined. As the fundamental
equation (4) is exact, any numerical deviation observed on the final
results are due to the particular measuring recipe one chooses to
adopt, not to the method itself. This is a big advantage, once one
can choose a more adequate measuring instrument for each different
system under study. The only requirement to obtain correct
microcanonical averages is a uniform probability distribution {\bf
inside each energy level, separately}. Detailed balance between {\bf
different energy levels} is irrelevant for BHM. {\bf All} possible
transitions between different energy levels are {\bf exactly} counted
by the BHM quantities $N^{\rm up}$ and $N^{\rm dn}$ themselves, {\bf
not} by the particular stochastic dynamic recipe one adopts. On the
other hand, besides the uniform sampling probability inside each
energy level, multicanonical methods also depend on the particular
transition rates between different energies, which are tuned during
the computer run in order to get a flat distribution of visits at the
end. Detailed balance is fundamental for multicanonical methods.
Thus, besides the acccuracy advantage commented before, any good
dynamic rule for multicanonical sampling is also good for BHM, but
the reverse is not true. Indeed, BHM allows the user to design his
own profile of visits along the energy axis, taking for instance a
better statistics near the critical region.

        During the past half century, theoretical studies provided us
with many recipes concerning detailed balance for Markovian
processes, leading to Gibbs equilibrium distribution. These recipes
show us how to design adequate transition rates between different
energy levels. Unfortunately, there are no equivalent theoretical
studies in what concerns adequate rules leading to a uniform
probability inside each energy level. Perhaps this missing point is
due to the fact that most studies were directed towards canonical
distributions, where a sharp region of the energy axis is visited:
after some transient steps (normally discarded from the Markovian
chain) the system becomes trapped into this sharp region. Thus, the
spread over this restricted region is supposed to occur naturally,
giving rise to the required uniformity inside each level. At least
for smooth $E$-dependent quantities $<Q(E)>$, the quoted sharpness
solves the problem. However, this is not the case if one needs to
cover wide energy ranges. For both multicanonical sampling as well as
BHM, this is an open problem waiting for new ideas, new insight. How
to assure a uniform probability of visits inside each energy level?
Some clues towards the answer were discussed, and some very efficient
practical solutions were proposed in section V.

\vskip 1cm
{\bf Acknowledgements}\parG

        This work was partially supported by brazilian agencies CAPES,
CNPq and FAPERJ. I am deeply indebted to Jian-Sheng Wang for exhaustive
and elucidative discussions about the subtleties inherent to this matter.
These discussions were, at least, very useful in showing me my ignorance
about the subject, and perhaps also showing us some possible ways to learn
a little bit. I am also indebted to my collaborators Hans Herrmann,
Thadeu Penna and mainly Adriano Lima, the youngest one who nevertheless
becames the most expert on BHM.

\vfill\eject
{\bf References}\parG

\item{[1]} P.M.C. de Oliveira, T.J.P. Penna and H.J. Herrmann, {\it Braz. 
J. Phys.} {\bf 26}, 677 (1996) (also in Cond-Mat 9610041).\par

\item{[2]} P.M.C. de Oliveira, {\it Eur. Phys. J.} {\bf B6}, 111 (1998) (also
in Cond-Mat 9807354).\par

\item{[3]} Z.W. Salzburg, J.D. Jacobson, W. Fickett and W.W. Wood,
{\it J. Chem. Phys.} {\bf 30}, 65 (1959); R.W. Swendsen, {\it Physica} {\bf A194}, 53
(1993), and references therein.\par

\item{[4]} G.M. Torrie and J.P.  Valleau, {\it Chem. Phys. Lett.}
{\bf 28}, 578 (1974); B. Bhanot, S.  Black, P. Carter and S.
Salvador, {\it Phys. Lett.} {\bf B183}, 381 (1987); M. Karliner, S.
Sharpe and Y. Chang, {\it Nucl. Phys.} {\bf B302}, 204 (1988).\par

\item{[5]} B.A. Berg and T. Neuhaus, {\it Phys. Lett.} {\bf B267},
249 (1991); A.P. Lyubartsev, A.A. Martsinovski, S.V. Shevkunov and
P.N. Vorontsov-Velyaminov, {\it J. Chem. Phys.} {\bf 96}, 1776
(1992); E. Marinari and G. Parisi, {\it Europhys. Lett.} {\bf 19},
451 (1992); B.A. Berg, {\it Int. J. Mod. Phys.} {\bf C4}, 249
(1993).\par

\item{[6]} J. Lee, {\it Phys. Rev. Lett.} {\bf 71}, 211 (1993).\par

\item{[7]} B. Hesselbo and R.B. Stinchcombe, {\it Phys. Rev. Lett.}
{\bf 74}, 2151 (1995).\par

\item{[8]} All real systems of interest are confined to a finite volume, thus
the energy spectrum will be always discrete. Nevertheless, the method can be
easily generalized also for the case of continuous energy spectra
models, as in\par
\item{} J.D. Mu\~noz and H.J. Herrmann, {\it Int. J. Mod. Phys.} {\bf
C10}, 95 (1999); {\it Comput. Phys. Comm.} {\bf 121-122}, 13 (1999).\par

\item{[9]} J.-S. Wang, T.K. Tay and R.H. Swendsen, {\it Phys. Rev.
Lett.} {\bf 82}, 476 (1999).\par

\item{[10]} G.R. Smith and A.D. Bruce, {\it J. Phys.} {\bf A28}, 6623
(1995); {\it Phys. Rev.} {\bf E53}, 6530 (1996); {\it Europhys.
Lett.} {\bf 39}, 91 (1996).\par

\item{[11]} J.-S. Wang and L.W. Lee, Cond-Mat 9903224.\par

\item{[12]} M. Kastner, J.D. Munoz and M. Promberger, Cond-Mat 9906097.\par

\item{[13]} R.H. Swendsen, J.-S. Wang, S.-T. Li, C. Genovese, B.
Diggs and J.B. Kadane, Cond-Mat 9908461.\par

\item{[14]} J.-S. Wang, {\it Comput. Phys. Comm.} {\bf 121-122},
22 (1999); Cond-Mat 9909177.\par

\item{[15]} A.R. de Lima, P.M.C. de Oliveira and T.J.P. Penna, {\it
Solid State Comm.} (2000) (also in Cond-Mat 9912152).\par

\item{[16]} Energy $E = 2$ is not allowed, because one cannot arrange the
spins on the square lattice in such a way that only two unsatisfied links
appear. The same occurs also for $E = 2L^2-2$. Besides these two cases, any
even value of $E$ in between 0 and $2L^2$ is allowed.\par

\item{[17]} P.M.C. de Oliveira, T.J.P. Penna and H.J. Herrmann, {\it Eur. 
Phys. J.} {\bf B1}, 205 (1998); P.M.C. de Oliveira, in {\sl Computer
Simulation Studies in Condensed Matter Physics} {\bf XI}, 169, eds. D.P.
Landau and H.-B. Sch\"uttler, Springer, Heidelberg/Berlin (1998).\par

\item{[18]} P.M.C. de Oliveira, {\it Int, J. Mod. Phys.} {\bf C9}, 497
(1998).\par

\item{[19]} P.M.C. de Oliveira, {\it Comput. Phys. Comm.} {\bf
121-122}, 16 (1999), presented at the APS/EPS Conference on
Computational Physics, Granada, Spain (1998).\par

\item{[20]} H.J. Herrmann, {\it J. Stat. Phys.} {\bf 45}, 145 (1986);
J.G. Zabolitzky and H.J. Herrmann, {\it J. Comp. Phys.} {\bf 76},
426 (1988), and references therein.\par

\item{[21]} S.C. Glotzer, D. Stauffer and S. Satry, {\it Physica} {\bf A164},
1 (1990).\par

\item{[22]} C. Moukarzel, {\it J. Phys.} {\bf A22}, 4487 (1989).\par

\item{[23]} M. Schulte, W. Stiefelhaben and E.S. Demme, {\it J. Phys.} {\bf
A20}, L1023 (1987).\par

\item{[24]} M. Creutz, {\it Phys. Rev. Lett.} {\bf 50}, 1411 (1983).\par

\item{[25]} C.M. Care, {\it J. Phys.} {\bf A29}, L505 (1996).\par

\item{[26]} B. Berg, {\it Nature} {\bf 361}, 708 (1993).\par

\item{[27]} J.-S. Wang, {\it Eur. Phys. J.} {\bf B8}, 287 (1999) (also
in Cond-Mat 9810017).\par

\item{[28]} B. Berg and U.H.E. Hansmann, {\it Eur. Phys. J.} {\bf
B6}, 395 (1998).\par

\item{[29]} A.R. de Lima, P.M.C. de Oliveira and T.J.P. Penna, {\it J.
Stat. Phys.} (2000) (also in Cond-Mat 0002176).\par

\item{[30]} J.L. Lebowitz, J.K. Percus and L. Verlet, {\it Phys.
Rev.} {\bf 153}, 250 (1967); D.H.E. Gross, {\it Phys. Rep.} {\bf
279}, 119 (1997), and references therein; D.H.E. Gross, Cond-Mat
9805391; M. Kastner, M. Promberger and A. H\"uller, in {\sl Computer
Simulation Studies in Condensed Matter Physics} {\bf XI}, eds. D.P.
Landau and H.-B. Sch\"uttler, Springer, Heidelberg/Berlin (1998); M.
Promberger, M. Kastner and A. H\"uller, Cond-Mat 9904265.\par

\item{[31]} N. Metropolis, A.W. Rosenbluth, M.N. Rosenbluth, A.H. Teller
and E. Teller, {\it J. Chem. Phys.} {\bf 21}, 1087 (1953).\par

\item{[32]} P.M.C. de Oliveira, T.J.P. Penna, S. Moss de Oliveira and
R.M. Zorzenon, {\it J. Phys.} {\bf A24}, 219 (1991); P.M.C. de Oliveira,
{\sl Computing Boolean Statistical Models}, World Scientific, Sigapore
London New York, ISBN 981-02-0238-5 (1991).\par

\item{[33]} P.D. Beale, {\it Phys. Rev. Lett.} {\bf 76}, 78 (1996).\par

\item{[34]} Surprisingly, probably due to some implementation
mistake, Table 2(b) of reference [27] presents wrong values for the
same data now correctly shown in Table IX. The reader could interpret
this as a failure of the flat histogram dynamics [27], but this is
not the case.\par

\vfill\eject
{\bf Tables}\parG

\parG\settabs 6\columns
\+$E$ \hfill $g(E)$\hskip 10pt&
\hfill $<N^{\rm up}(E)>$ \hfill&
\hfill $<N^{\rm up}(E)>$ \hfill&
\hfill $<N^{\rm tie}(E)>$ \hfill&
\hfill $<N^{\rm dn}(E)>$\hfill&
\hfill $<N^{\rm dn}(E)>$\hfill&\cr

\+ &
\hfill $(\Delta{E}=4)$ \hfill&
\hfill $(\Delta{E}=2)$ \hfill&
\hfill $(\Delta{E}=0)$ \hfill&
\hfill $(\Delta{E}=2)$\hfill&
\hfill $(\Delta{E}=4)$\hfill&\cr

\+\hskip 5pt$0$ \hfill $2$\hskip 10pt&
\hfill $16$ \hfill&
\hfill $0$ \hfill&
\hfill $0$ \hfill&
\hfill $0$ \hfill&
\hfill $0$ \hfill&\cr

\+\hskip 5pt$4$ \hfill $32$\hskip 10pt&
\hfill $11$ \hfill&
\hfill $4$ \hfill&
\hfill $0$ \hfill&
\hfill $0$ \hfill&
\hfill $1$ \hfill&\cr

\+\hskip 5pt$6$ \hfill $64$\hskip 10pt&
\hfill $8$ \hfill&
\hfill $6$ \hfill&
\hfill $0$ \hfill&
\hfill $2$ \hfill&
\hfill $0$ \hfill&\cr

\+\hskip 5pt$8$ \hfill $424$\hskip 10pt&
\hfill $6.11321$ \hfill&
\hfill $6.33962$ \hfill&
\hfill $1.81132$ \hfill&
\hfill $0.90566$ \hfill&
\hfill $0.83019$ \hfill&\cr

\+$10$ \hfill $1728$\hskip 10pt&
\hfill $3.55556$ \hfill&
\hfill $7.03704$ \hfill&
\hfill $3.55556$ \hfill&
\hfill $1.55556$ \hfill&
\hfill $0.29630$ \hfill&\cr

\+$12$ \hfill $6688$\hskip 10pt&
\hfill $2.31101$ \hfill&
\hfill $5.97129$ \hfill&
\hfill $5.51196$ \hfill&
\hfill $1.81818$ \hfill&
\hfill $0.38756$ \hfill&\cr

\+$14$ \hfill $13568$\hskip 10pt&
\hfill $1.13208$ \hfill&
\hfill $5.58491$ \hfill&
\hfill $5.88679$ \hfill&
\hfill $2.94340$ \hfill&
\hfill $0.45283$ \hfill&\cr

\+$16$ \hfill $20524$\hskip 10pt&
\hfill $0.75307$ \hfill&
\hfill $3.69207$ \hfill&
\hfill $7.10973$ \hfill&
\hfill $3.69207$ \hfill&
\hfill $0.75307$ \hfill&\cr\parG

\item{Table I} Exact data concerning the $4 \times 4$ square lattice
Ising ferromagnet. The second half of the spectrum ($E = 18
\dots 32$) is symmetric the to first half ($E = 14 \dots 0$). $N^{\rm
tie}$ is the number of potential movements (one-spin flips) keeping
the energy unchanged.\par
\vskip 1cm

\parG\settabs 6\columns
\+$E$ \hfill $V(E)$\hskip 10pt&
\hfill $<N^{\rm up}(E)>$ \hfill&
\hfill $<N^{\rm up}(E)>$ \hfill&
\hfill $<N^{\rm tie}(E)>$ \hfill&
\hfill $<N^{\rm dn}(E)>$\hfill&
\hfill $<N^{\rm dn}(E)>$\hfill&\cr

\+\hfill $\times 10^8$\hskip 10pt&
\hfill $(\Delta{E}=4)$ \hfill&
\hfill $(\Delta{E}=2)$ \hfill&
\hfill $(\Delta{E}=0)$ \hfill&
\hfill $(\Delta{E}=2)$\hfill&
\hfill $(\Delta{E}=4)$\hfill&\cr

\+\hskip 5pt$8$ \hfill $5.62$\hskip 10pt&
\hfill $6.1145$ \hfill&
\hfill $6.3376$ \hfill&
\hfill $1.81155$ \hfill&
\hfill $0.90606$ \hfill&
\hfill $0.83028$ \hfill&\cr

\+$10$ \hfill $6.11$\hskip 10pt&
\hfill $3.5565$ \hfill&
\hfill $7.0366$ \hfill&
\hfill $3.5540$ \hfill&
\hfill $1.55630$ \hfill&
\hfill $0.29665$ \hfill&\cr

\+$12$ \hfill $6.32$\hskip 10pt&
\hfill $2.3111$ \hfill&
\hfill $5.9711$ \hfill&
\hfill $5.5119$ \hfill&
\hfill $1.81843$ \hfill&
\hfill $0.38744$ \hfill&\cr

\+$14$ \hfill $3.42$\hskip 10pt&
\hfill $1.13151$ \hfill&
\hfill $5.5851$ \hfill&
\hfill $5.8879$ \hfill&
\hfill $2.9429$ \hfill&
\hfill $0.45259$ \hfill&\cr

\+$16$ \hfill $1.38$\hskip 10pt&
\hfill $0.75302$ \hfill&
\hfill $3.6916$ \hfill&
\hfill $7.1109$ \hfill&
\hfill $3.6913$ \hfill&
\hfill $0.75315$ \hfill&\cr\parG

\item{Table II} Canonical simulation for the $4 \times 4$ square
lattice Ising ferromagnet, with a fixed temperature $T(8) =
3.02866216$ for which the exact energy average is $<E>_T\, = 8$. For
all averaged values, statistical fluctuations fall on the two last
digits, at most.\par
\parG

\vfill\eject

\parG\settabs 6\columns
\+$E$ \hfill $V(E)$\hskip 10pt&
\hfill $<N^{\rm up}(E)>$ \hfill&
\hfill $<N^{\rm up}(E)>$ \hfill&
\hfill $<N^{\rm tie}(E)>$ \hfill&
\hfill $<N^{\rm dn}(E)>$\hfill&
\hfill $<N^{\rm dn}(E)>$\hfill&\cr

\+\hfill $\times 10^8$\hskip 10pt&
\hfill $(\Delta{E}=4)$ \hfill&
\hfill $(\Delta{E}=2)$ \hfill&
\hfill $(\Delta{E}=0)$ \hfill&
\hfill $(\Delta{E}=2)$\hfill&
\hfill $(\Delta{E}=4)$\hfill&\cr

\+\hskip 5pt$8$ \hfill $4.68$\hskip 10pt&
\hfill $6.1123$ \hfill&
\hfill $6.3410$ \hfill&
\hfill $1.81107$ \hfill&
\hfill $0.90577$ \hfill&
\hfill $0.82989$ \hfill&\cr

\+$10$ \hfill $6.22$\hskip 10pt&
\hfill $3.5559$ \hfill&
\hfill $7.0367$ \hfill&
\hfill $3.5552$ \hfill&
\hfill $1.55578$ \hfill&
\hfill $0.29639$ \hfill&\cr

\+$12$ \hfill $7.85$\hskip 10pt&
\hfill $2.3113$ \hfill&
\hfill $5.9713$ \hfill&
\hfill $5.5115$ \hfill&
\hfill $1.81836$ \hfill&
\hfill $0.38771$ \hfill&\cr

\+$14$ \hfill $5.20$\hskip 10pt&
\hfill $1.13148$ \hfill&
\hfill $5.5858$ \hfill&
\hfill $5.8870$ \hfill&
\hfill $2.9426$ \hfill&
\hfill $0.45307$ \hfill&\cr

\+$16$ \hfill $2.57$\hskip 10pt&
\hfill $0.75274$ \hfill&
\hfill $3.6914$ \hfill&
\hfill $7.1116$ \hfill&
\hfill $3.6917$ \hfill&
\hfill $0.75257$ \hfill&\cr\parG

\item{Table III} The same as Table II, now with a fixed temperature
$T(10) = 3.57199419$ for which the exact energy average is $<E>_T\, =
10$.\par
\vskip 1cm

\parG\settabs 6\columns
\+$E$ \hfill $V(E)$\hskip 10pt&
\hfill $<N^{\rm up}(E)>$ \hfill&
\hfill $<N^{\rm up}(E)>$ \hfill&
\hfill $<N^{\rm tie}(E)>$ \hfill&
\hfill $<N^{\rm dn}(E)>$\hfill&
\hfill $<N^{\rm dn}(E)>$\hfill&\cr

\+\hfill $\times 10^8$\hskip 10pt&
\hfill $(\Delta{E}=4)$ \hfill&
\hfill $(\Delta{E}=2)$ \hfill&
\hfill $(\Delta{E}=0)$ \hfill&
\hfill $(\Delta{E}=2)$\hfill&
\hfill $(\Delta{E}=4)$\hfill&\cr

\+\hskip 5pt$8$ \hfill $3.08$\hskip 10pt&
\hfill $6.1122$ \hfill&
\hfill $6.3404$ \hfill&
\hfill $1.8116$ \hfill&
\hfill $0.90667$ \hfill&
\hfill $0.82909$ \hfill&\cr

\+$10$ \hfill $5.34$\hskip 10pt&
\hfill $3.5550$ \hfill&
\hfill $7.0372$ \hfill&
\hfill $3.5552$ \hfill&
\hfill $1.55529$ \hfill&
\hfill $0.29597$ \hfill&\cr

\+$12$ \hfill $8.78$\hskip 10pt&
\hfill $2.31065$ \hfill&
\hfill $5.9707$ \hfill&
\hfill $5.5138$ \hfill&
\hfill $1.81777$ \hfill&
\hfill $0.38711$ \hfill&\cr

\+$14$ \hfill $7.56$\hskip 10pt&
\hfill $1.13206$ \hfill&
\hfill $5.5846$ \hfill&
\hfill $5.8872$ \hfill&
\hfill $2.9436$ \hfill&
\hfill $0.45255$ \hfill&\cr

\+$16$ \hfill $4.85$\hskip 10pt&
\hfill $0.75310$ \hfill&
\hfill $3.6923$ \hfill&
\hfill $7.1091$ \hfill&
\hfill $3.6924$ \hfill&
\hfill $0.75303$ \hfill&\cr\parG

\item{Table IV} The same again, now with a fixed temperature $T(12) =
4.66862103$ for which the exact energy average is $<E>_T\, = 12$.\par
\vskip 1cm

\parG\settabs 6\columns
\+$E$ \hfill $V(E)$\hskip 10pt&
\hfill $<N^{\rm up}(E)>$ \hfill&
\hfill $<N^{\rm up}(E)>$ \hfill&
\hfill $<N^{\rm tie}(E)>$ \hfill&
\hfill $<N^{\rm dn}(E)>$\hfill&
\hfill $<N^{\rm dn}(E)>$\hfill&\cr

\+\hfill $\times 10^8$\hskip 10pt&
\hfill $(\Delta{E}=4)$ \hfill&
\hfill $(\Delta{E}=2)$ \hfill&
\hfill $(\Delta{E}=0)$ \hfill&
\hfill $(\Delta{E}=2)$\hfill&
\hfill $(\Delta{E}=4)$\hfill&\cr

\+\hskip 5pt$8$ \hfill $1.30$\hskip 10pt&
\hfill $6.1137$ \hfill&
\hfill $6.3404$ \hfill&
\hfill $1.8098$ \hfill&
\hfill $0.90444$ \hfill&
\hfill $0.83169$ \hfill&\cr

\+$10$ \hfill $3.27$\hskip 10pt&
\hfill $3.5550$ \hfill&
\hfill $7.0379$ \hfill&
\hfill $3.5554$ \hfill&
\hfill $1.55540$ \hfill&
\hfill $0.29628$ \hfill&\cr

\+$12$ \hfill $7.84$\hskip 10pt&
\hfill $2.31063$ \hfill&
\hfill $5.9716$ \hfill&
\hfill $5.5123$ \hfill&
\hfill $1.81816$ \hfill&
\hfill $0.38735$ \hfill&\cr

\+$14$ \hfill $9.83$\hskip 10pt&
\hfill $1.13243$ \hfill&
\hfill $5.5845$ \hfill&
\hfill $5.8866$ \hfill&
\hfill $2.9436$ \hfill&
\hfill $0.45291$ \hfill&\cr

\+$16$ \hfill $9.20$\hskip 10pt&
\hfill $0.75357$ \hfill&
\hfill $3.6916$ \hfill&
\hfill $7.1094$ \hfill&
\hfill $3.6920$ \hfill&
\hfill $0.75339$ \hfill&\cr\parG

\item{Table V} The same once more, now with a fixed temperature
$T(14) = 8.33883787$ for which the exact energy average is $<E>_T\, =
14$.\par
\parG

\vfill\eject

\parG\settabs 6\columns
\+$E$ \hfill $V(E)$\hskip 10pt&
\hfill $<N^{\rm up}(E)>$ \hfill&
\hfill $<N^{\rm up}(E)>$ \hfill&
\hfill $<N^{\rm tie}(E)>$ \hfill&
\hfill $<N^{\rm dn}(E)>$\hfill&
\hfill $<N^{\rm dn}(E)>$\hfill&\cr

\+\hfill $\times 10^8$\hskip 10pt&
\hfill $(\Delta{E}=4)$ \hfill&
\hfill $(\Delta{E}=2)$ \hfill&
\hfill $(\Delta{E}=0)$ \hfill&
\hfill $(\Delta{E}=2)$\hfill&
\hfill $(\Delta{E}=4)$\hfill&\cr

\+\hskip 5pt$8$ \hfill $0.35$\hskip 10pt&
\hfill $6.1132$ \hfill&
\hfill $6.3405$ \hfill&
\hfill $1.8106$ \hfill&
\hfill $0.90464$ \hfill&
\hfill $0.83108$ \hfill&\cr

\+$10$ \hfill $1.40$\hskip 10pt&
\hfill $3.5560$ \hfill&
\hfill $7.0366$ \hfill&
\hfill $3.5552$ \hfill&
\hfill $1.55577$ \hfill&
\hfill $0.29642$ \hfill&\cr

\+$12$ \hfill $5.22$\hskip 10pt&
\hfill $2.31103$ \hfill&
\hfill $5.9717$ \hfill&
\hfill $5.5112$ \hfill&
\hfill $1.81838$ \hfill&
\hfill $0.38769$ \hfill&\cr

\+$14$ \hfill $10.17$\hskip 10pt&
\hfill $1.13244$ \hfill&
\hfill $5.5847$ \hfill&
\hfill $5.8863$ \hfill&
\hfill $2.9436$ \hfill&
\hfill $0.45298$ \hfill&\cr

\+$16$ \hfill $14.77$\hskip 10pt&
\hfill $0.75265$ \hfill&
\hfill $3.6922$ \hfill&
\hfill $7.1102$ \hfill&
\hfill $3.6922$ \hfill&
\hfill $0.75266$ \hfill&\cr\parG

\item{Table VI} Again, now with a fixed temperature $T = 100$,
mimicking $T(16) = \infty$.\par
\vskip 1cm

\parG\settabs 6\columns
\+$E$ \hfill $V(E)$\hskip 10pt&
\hfill $<N^{\rm up}(E)>$ \hfill&
\hfill $<N^{\rm up}(E)>$ \hfill&
\hfill $<N^{\rm tie}(E)>$ \hfill&
\hfill $<N^{\rm dn}(E)>$\hfill&
\hfill $<N^{\rm dn}(E)>$\hfill&\cr

\+\hfill $\times 10^8$\hskip 10pt&
\hfill $(\Delta{E}=4)$ \hfill&
\hfill $(\Delta{E}=2)$ \hfill&
\hfill $(\Delta{E}=0)$ \hfill&
\hfill $(\Delta{E}=2)$\hfill&
\hfill $(\Delta{E}=4)$\hfill&\cr

\+\hskip 5pt$8$ \hfill $3.91$\hskip 10pt&
\hfill $6.1075$ \hfill&
\hfill $6.3168$ \hfill&
\hfill $1.8413$ \hfill&
\hfill $0.93716$ \hfill&
\hfill $0.79731$ \hfill&\cr

\+$10$ \hfill $4.11$\hskip 10pt&
\hfill $3.5647$ \hfill&
\hfill $7.0105$ \hfill&
\hfill $3.5752$ \hfill&
\hfill $1.55947$ \hfill&
\hfill $0.29019$ \hfill&\cr

\+$12$ \hfill $4.96$\hskip 10pt&
\hfill $2.32037$ \hfill&
\hfill $5.9336$ \hfill&
\hfill $5.5420$ \hfill&
\hfill $1.83372$ \hfill&
\hfill $0.37031$ \hfill&\cr

\+$14$ \hfill $3.99$\hskip 10pt&
\hfill $1.13602$ \hfill&
\hfill $5.5671$ \hfill&
\hfill $5.8935$ \hfill&
\hfill $2.9675$ \hfill&
\hfill $0.43584$ \hfill&\cr

\+$16$ \hfill $3.55$\hskip 10pt&
\hfill $0.73739$ \hfill&
\hfill $3.6917$ \hfill&
\hfill $7.1232$ \hfill&
\hfill $3.7290$ \hfill&
\hfill $0.71873$ \hfill&\cr\parG

\item{Table VII} Random walk, pseudo-canonical simulation with
asymmetric $E$-dependent Boltzmann acceptance probability
$\exp[-\Delta{E}/T(E)]$, to be compared with the exact values in
Table I.\par
\vskip 1cm

\parG\settabs 6\columns
\+$E$ \hfill $V(E)$\hskip 10pt&
\hfill $<N^{\rm up}(E)>$ \hfill&
\hfill $<N^{\rm up}(E)>$ \hfill&
\hfill $<N^{\rm tie}(E)>$ \hfill&
\hfill $<N^{\rm dn}(E)>$\hfill&
\hfill $<N^{\rm dn}(E)>$\hfill&\cr

\+\hfill $\times 10^8$\hskip 10pt&
\hfill $(\Delta{E}=4)$ \hfill&
\hfill $(\Delta{E}=2)$ \hfill&
\hfill $(\Delta{E}=0)$ \hfill&
\hfill $(\Delta{E}=2)$\hfill&
\hfill $(\Delta{E}=4)$\hfill&\cr

\+\hskip 5pt$8$ \hfill $3.25$\hskip 10pt&
\hfill $6.1134$ \hfill&
\hfill $6.3392$ \hfill&
\hfill $1.81132$ \hfill&
\hfill $0.90608$ \hfill&
\hfill $0.82998$ \hfill&\cr

\+$10$ \hfill $3.87$\hskip 10pt&
\hfill $3.5571$ \hfill&
\hfill $7.0343$ \hfill&
\hfill $3.5561$ \hfill&
\hfill $1.55637$ \hfill&
\hfill $0.29609$ \hfill&\cr

\+$12$ \hfill $5.48$\hskip 10pt&
\hfill $2.31315$ \hfill&
\hfill $5.9677$ \hfill&
\hfill $5.5123$ \hfill&
\hfill $1.81968$ \hfill&
\hfill $0.38715$ \hfill&\cr

\+$14$ \hfill $5.57$\hskip 10pt&
\hfill $1.13256$ \hfill&
\hfill $5.5833$ \hfill&
\hfill $5.8871$ \hfill&
\hfill $2.9458$ \hfill&
\hfill $0.45130$ \hfill&\cr

\+$16$ \hfill $6.56$\hskip 10pt&
\hfill $0.75220$ \hfill&
\hfill $3.6918$ \hfill&
\hfill $7.1109$ \hfill&
\hfill $3.6939$ \hfill&
\hfill $0.75118$ \hfill&\cr\parG

\item{Table VIII} Random walk, pseudo-canonical simulation with
symmetric $E$-dependent Boltzmann acceptance probability
$\exp[-\Delta{E}/T(E+\Delta{E}/2)]$, to be compared with the exact
values in Table I.\par

\vfill\eject

\parG\settabs 6\columns
\+$E$ \hfill $V(E)$\hskip 10pt&
\hfill $<N^{\rm up}(E)>$ \hfill&
\hfill $<N^{\rm up}(E)>$ \hfill&
\hfill $<N^{\rm tie}(E)>$ \hfill&
\hfill $<N^{\rm dn}(E)>$\hfill&
\hfill $<N^{\rm dn}(E)>$\hfill&\cr

\+\hfill $\times 10^8$\hskip 10pt&
\hfill $(\Delta{E}=4)$ \hfill&
\hfill $(\Delta{E}=2)$ \hfill&
\hfill $(\Delta{E}=0)$ \hfill&
\hfill $(\Delta{E}=2)$\hfill&
\hfill $(\Delta{E}=4)$\hfill&\cr

\+\hskip 5pt$8$ \hfill $4.53$\hskip 10pt&
\hfill $6.1123$ \hfill&
\hfill $6.3407$ \hfill&
\hfill $1.81151$ \hfill&
\hfill $0.90579$ \hfill&
\hfill $0.82973$ \hfill&\cr

\+$10$ \hfill $4.53$\hskip 10pt&
\hfill $3.5553$ \hfill&
\hfill $7.0376$ \hfill& 
\hfill $3.5552$ \hfill&
\hfill $1.55561$ \hfill&
\hfill $0.29630$ \hfill&\cr

\+$12$ \hfill $4.53$\hskip 10pt&
\hfill $2.3112$ \hfill&
\hfill $5.9704$ \hfill&
\hfill $5.5130$ \hfill&
\hfill $1.81807$ \hfill&
\hfill $0.38733$ \hfill&\cr

\+$14$ \hfill $4.53$\hskip 10pt&
\hfill $1.13190$ \hfill&
\hfill $5.5848$ \hfill&
\hfill $5.8873$ \hfill&
\hfill $2.9432$ \hfill&
\hfill $0.45272$ \hfill&\cr

\+$16$ \hfill $4.53$\hskip 10pt&
\hfill $0.75283$ \hfill&
\hfill $3.6920$ \hfill&
\hfill $7.1108$ \hfill&
\hfill $3.6912$ \hfill&
\hfill $0.75321$ \hfill&\cr\parG

\item{Table IX} Flat histogram [27] simulation with
symmetric $E$-dependent Boltzmann acceptance probability
$\exp[-\Delta{E}/T_m(E)]$, equation (9), to be compared with the exact
values in Table I.\par

\vskip 1cm

\parG\settabs 6\columns
\+$E$ \hfill $V(E)$\hskip 10pt&
\hfill $<N^{\rm up}(E)>$ \hfill&
\hfill $<N^{\rm up}(E)>$ \hfill&
\hfill $<N^{\rm tie}(E)>$ \hfill&
\hfill $<N^{\rm dn}(E)>$\hfill&
\hfill $<N^{\rm dn}(E)>$\hfill&\cr

\+\hfill $\times 10^8$\hskip 10pt&
\hfill $(\Delta{E}=4)$ \hfill&
\hfill $(\Delta{E}=2)$ \hfill&
\hfill $(\Delta{E}=0)$ \hfill&
\hfill $(\Delta{E}=2)$\hfill&
\hfill $(\Delta{E}=4)$\hfill&\cr

\+\hskip 5pt$8$ \hfill $3.51$\hskip 10pt&
\hfill $6.1131$ \hfill&
\hfill $6.3399$ \hfill&
\hfill $1.81121$ \hfill&
\hfill $0.90558$ \hfill&
\hfill $0.83023$ \hfill&\cr

\+$10$ \hfill $4.65$\hskip 10pt&
\hfill $3.5553$ \hfill&
\hfill $7.0372$ \hfill&
\hfill $3.5561$ \hfill&
\hfill $1.55532$ \hfill&
\hfill $0.29619$ \hfill&\cr

\+$12$ \hfill $17.03$\hskip 10pt&
\hfill $2.3106$ \hfill&
\hfill $5.9717$ \hfill&
\hfill $5.5124$ \hfill&
\hfill $1.81788$ \hfill&
\hfill $0.38749$ \hfill&\cr

\+$14$ \hfill $3.81$\hskip 10pt&
\hfill $1.13212$ \hfill&
\hfill $5.5850$ \hfill&
\hfill $5.8865$ \hfill&
\hfill $2.9434$ \hfill&
\hfill $0.45296$ \hfill&\cr

\+$16$ \hfill $2.13$\hskip 10pt&
\hfill $0.75298$ \hfill&
\hfill $3.6917$ \hfill&
\hfill $7.1108$ \hfill&
\hfill $3.6915$ \hfill&
\hfill $0.75305$ \hfill&\cr\parG

\item{Table X} Random walk dynamics [1] simulation with approximation
A1 (see text), to be compared with the exact values in Table I.\par

\vfill\eject

\parG\settabs 5\columns
\+  $E$& \hskip-30pt $\ln {g(E+2)\over g(E)}$
&&\hskip-10pt $\ln {<N_{\rm up}(E)>\over <N_{\rm dn}(E+2)>}$ &\cr
\vskip 10pt
\+ 300& \hskip-30pt 1.749921&1.749897&1.749800&1.749992\cr
\+ 302& \hskip-30pt 1.748650&1.748599&1.748496&1.748678\cr
\+ 304& \hskip-30pt 1.747398&1.747387&1.747288&1.747365\cr
\+ 306& \hskip-30pt 1.746164&1.746151&1.746055&1.746144\cr
\+ 308& \hskip-30pt 1.744946&1.744960&1.744834&1.744905\cr
\+ 310& \hskip-30pt 1.743743&1.743748&1.743638&1.743660\cr
\+ 312& \hskip-30pt 1.742554&1.742590&1.742452&1.742438\cr
\+ 314& \hskip-30pt 1.741375&1.741424&1.741305&1.741192\cr
\+ 316& \hskip-30pt 1.740206&1.740295&1.740148&1.740000\cr
\+ 318& \hskip-30pt 1.739045&1.739128&1.738995&1.738762\cr
\+ 320& \hskip-30pt 1.737889&1.737970&1.737828&1.737596\cr
\+ 322& \hskip-30pt 1.736737&1.736831&1.736655&1.736406\cr
\+ 324& \hskip-30pt 1.735586&1.735680&1.735494&1.735275\cr
\+ 326& \hskip-30pt 1.734434&1.734560&1.734342&1.734086\cr
\+ 328& \hskip-30pt 1.733279&1.733420&1.733174&1.732904\cr\parG

\item{Table XI} Data for the $32 \times 32$ square lattice, near the
critical region. Exact values are in the second column, and the next 2
columns correspond to different BHM runs. The random walk dynamics [1]
with approximation A1 (see text) was adopted. A new averaging state is
counted after each single-spin flip, instead of waiting for a whole
lattice sweep. Nearly $10^9$ states per energy were sampled for ech run.
The same numerical accuracy was obtained also by using the further
approximations A2 and A3, last column, in spite of the detailed balance
violation [27] between different energies, forced by A2. Within BHM, one
does not need to care about the relative balance of visitations to
different energies. Only a uniform sampling probability inside each energy
level, separately, is important.\parG

\vfill\eject
\parG{\bf Figure Captions}\parG

\item{Figure 1} Exact canonical temperature as a function of the
energy (continuous line), for a $4 \times 4$ square lattice Ising
ferromagnet. Symbols correspond to the microcanonical version of the
temperature, equation (8a), with energy gaps $\Delta{E} = 2$
(circles) or 4 (diamonds).\par

\item{Figure 2} The same as figure 1, now for a $32 \times 32$
lattice, with the same symbols, and equation (8b). Note that the
deviations between the true canonical temperatures and the
microcanonical versions are now restricted to the very beginning of
the energy spectrum (upper inset). At the critical region these
deviations are much smaller (lower inset, with a 100 times finer
scale). The deviations become smaller yet for larger systems.\par

\item{Figure 3} Specific heat for the $32 \times 32$ lattice Ising
ferromagnet, adopting the random walk dynamics within restricted
energy windows [19]. Circles show the exact values [33].\par

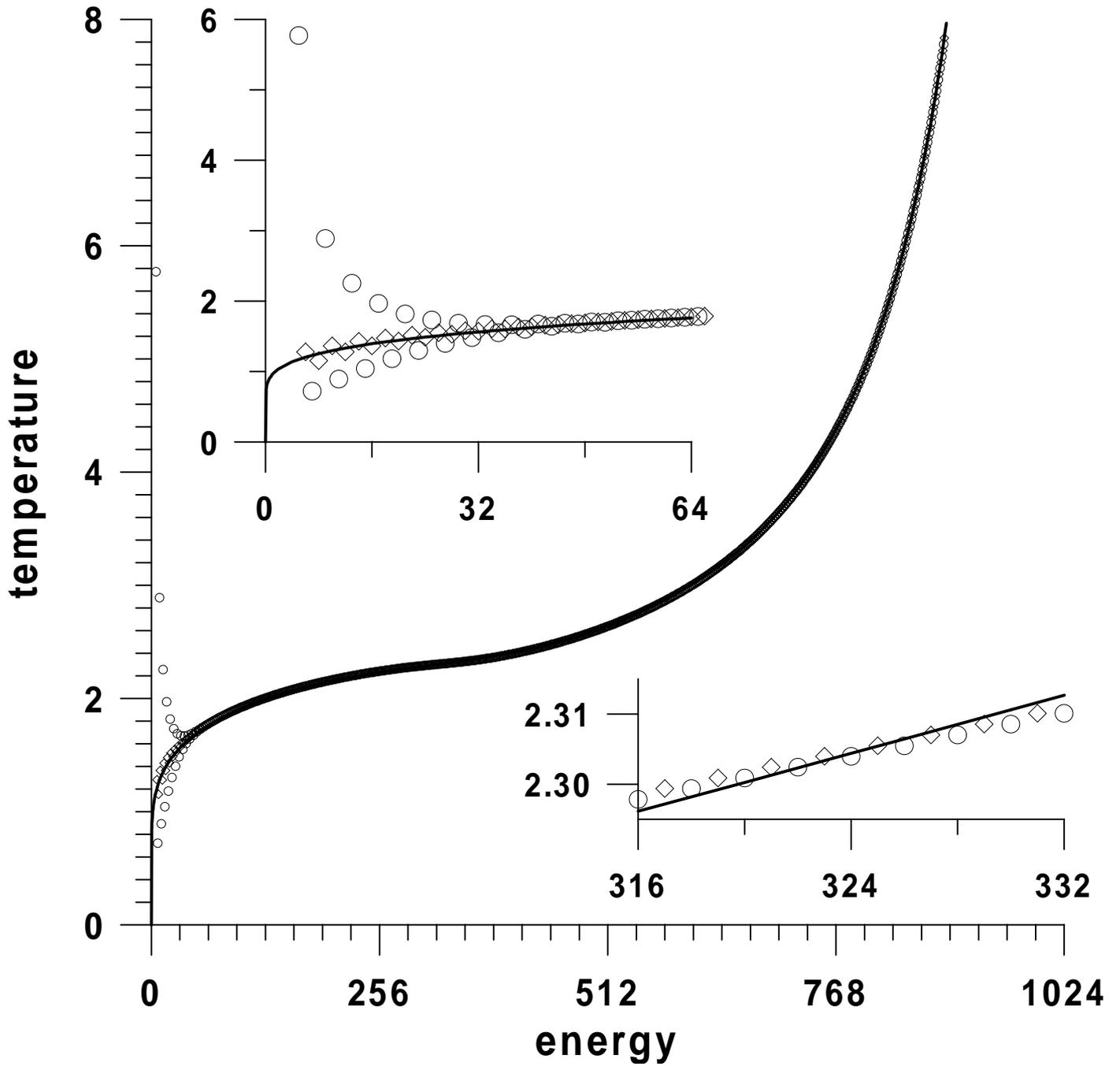
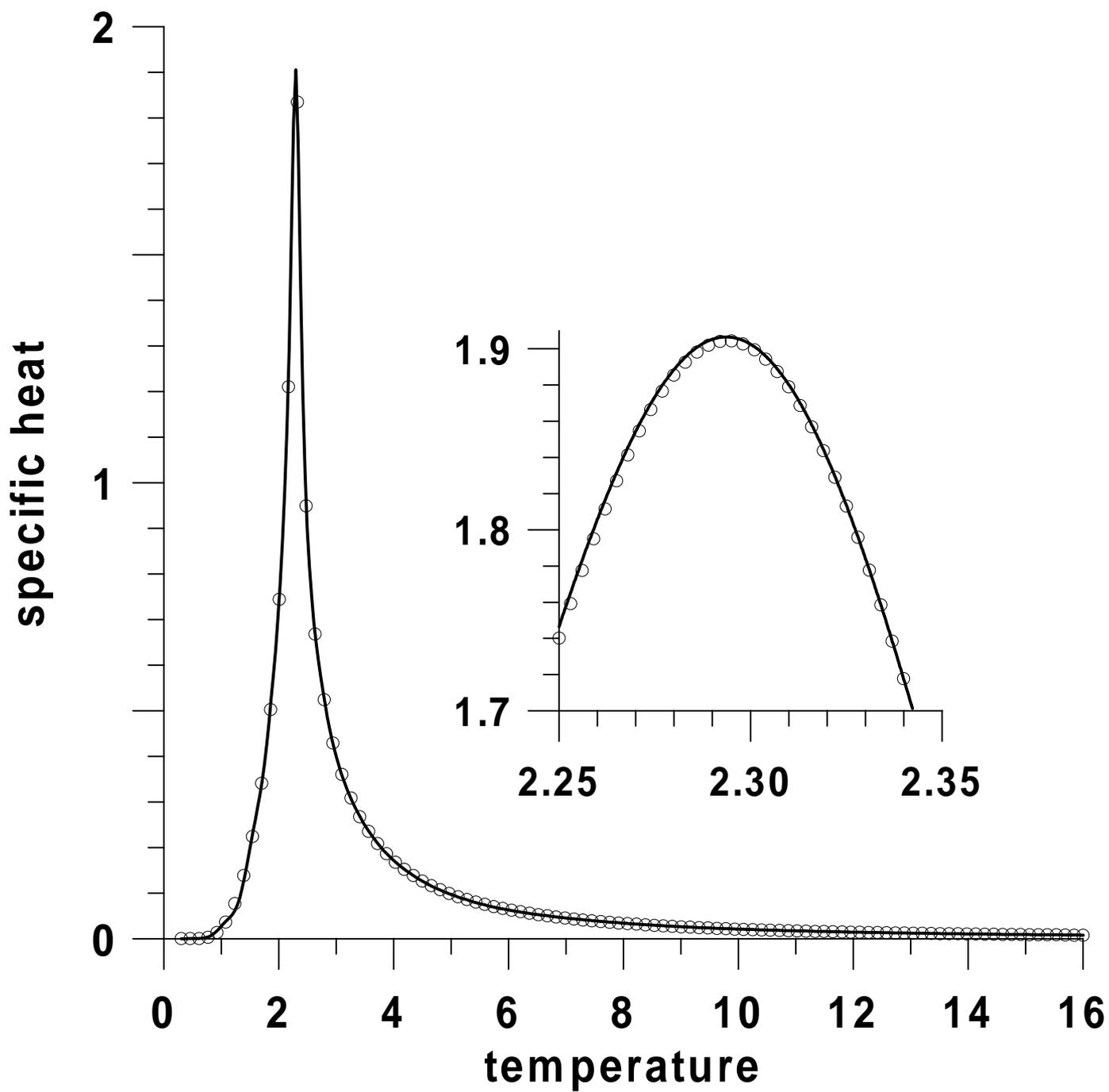
\bye